\documentclass{aa} 
\usepackage{latexsym}
\usepackage{amssymb}
\usepackage{graphicx}
\usepackage{natbib}
\usepackage{psfig}
\usepackage{txfonts}
\newcommand{\hii}{H~{\sc ii}}
\newcommand{\ha}{\ifmmode {\rm H}\alpha \else H$\alpha$\fi}
\newcommand{\hb}{\ifmmode {\rm H}\beta \else H$\beta$\fi}
\newcommand{\lya}{\ifmmode {\rm Ly}\alpha \else Ly$\alpha$\fi}

\newcommand{\heii}{He~{\sc ii}}
\newcommand{\Heiiuv}{He~{\sc ii} $\lambda$1640}

\newcommand{\qh}{\ifmmode q({\rm H}) \else $q({\rm H})$\fi}
\newcommand{\qhe}{\ifmmode q({\rm He^0}) \else $q({\rm He^0})$\fi}
\newcommand{\qhep}{\ifmmode q({\rm He^+}) \else $q({\rm He^+})$\fi}
\newcommand{\Qh}{\ifmmode Q({\rm H}) \else $Q({\rm H})$\fi}
\newcommand{\Qhe}{\ifmmode Q({\rm He^0}) \else $Q({\rm He^0})$\fi}
\newcommand{\Qhep}{\ifmmode Q({\rm He^+}) \else $Q({\rm He^+})$\fi}
\newcommand{\Qhtwo}{\ifmmode Q({\rm LW}) \else $Q({\rm LW})$\fi}
\newcommand{\qrathe}{\ifmmode q({\rm He^0})/q({\rm H}) \else $q({\rm He^0})/q({\rm H})$\fi}
\newcommand{\qrathep}{\ifmmode q({\rm He^+})/q({\rm H}) \else $q({\rm He^+})/q({\rm H})$\fi}
\newcommand{\Qrathe}{\ifmmode Q({\rm He^0})/Q({\rm H}) \else $Q({\rm He^0})/Q({\rm H})$\fi}
\newcommand{\Qrathep}{\ifmmode Q({\rm He^+})/Q({\rm H}) \else $Q({\rm He^+})/Q({\rm H})$\fi}
\newcommand{\tbb}{\ifmmode T_{\rm bb} \else $T_{\rm bb}$ \fi}
\newcommand{\Qhave}{\ifmmode \overline{Q}({\rm H}) \else $\overline{Q}({\rm H})$\fi}
\newcommand{\Qheave}{\ifmmode \overline{Q}({\rm He^0}) \else $\overline{Q}({\rm He^0})$\fi}
\newcommand{\Qhepave}{\ifmmode \overline{Q}({\rm He^+}) \else $\overline{Q}({\rm He^+})$\fi}
\newcommand{\Qhtwoave}{\ifmmode \overline{Q}({\rm H}_2) \else $\overline{Q}({\rm H}_2)$\fi}
\newcommand{\Qratheave}{\ifmmode \overline{Q}({\rm He^0})/\overline{Q}({\rm H}) \else $\overline{Q}({\rm He^0})/\\
overline{Q}({\rm H})$\fi}
\newcommand{\Qrathepave}{\ifmmode \overline{Q}({\rm He^+})/\overline{Q}({\rm H}) \else $\overline{Q}({\rm He^+})/\
\overline{Q}({\rm H})$\fi}
\newcommand{\zcoll}{\ifmmode Z_{\rm coll} \else $Z_{\rm coll}$\fi}

\def\mup{\ifmmode M_{\rm up} \else M$_{\rm up}$\fi}
\def\mlow{\ifmmode M_{\rm low} \else M$_{\rm low}$\fi}

\def\cmc{cm$^{-3}$}

\def\ergs{erg s$^{-1}$}

\def\msun{\ifmmode M_{\odot} \else M$_{\odot}$\fi}
\def\msunyr{\ifmmode M_{\odot} {\rm yr}^{-1} \else M$_{\odot}$ yr$^{-1}$\fi}
\def\zsun{\ifmmode Z_{\odot} \else Z$_{\odot}$\fi}
\def\zsun{\ifmmode Z_{\odot} \else Z$_{\odot}$\fi}
\def\lsun{\ifmmode L_{\odot} \else L$_{\odot}$\fi}
\begin{document}
\title{Predicted UV properties of very metal-poor starburst galaxies}

\authorrunning{A. Raiter, D. Schaerer, R. Fosbury}
\titlerunning{UV properties of starburst galaxies}

   \author{Anna Raiter\inst{1}, Daniel Schaerer \inst{2,3}
          \and
	 Robert A.E. Fosbury \inst{4} 
          }
   \offprints{A. Raiter}

   \institute{European Southern Observatory, Karl-Schwarzschild-Strasse 2,
Garching bei M\"unchen 85748, Germany\\  \email{araiter@eso.org}
   \and Geneva Observatory, University of Geneva,
              51, chemin des Maillettes, CH-1290 Versoix, Switzerland\\
              \email{daniel.schaerer@unige.ch}
         \and
             Observatoire Midi-Pyr\'{e}n\'{e}es, Laboratoire
             d`Astrophysique, UMR 5572, 14 Avenue E.Belin, F-31400
             Toulouse, France
             \and
             ST-ECF, Karl-Schwarzschild Str. 2, Garching bei M\"unchen 85748, Germany\\
             \email{rfosbury@eso.org}
             }

\date{Received date / Accepted date}

  \abstract
   {}
   {We study the expected properties of starburst galaxies in order to provide the point of reference for interpretation
   of high-$z$ galaxy surveys and of very metal-poor galaxies. 
   We concentrate mainly on the UV characteristics such as the ionizing spectra, the UV continuum,
   the \lya\ and \Heiiuv\ line and two-photon continuum emission.}
   {We use evolutionary synthesis models covering metallicities from Pop~III to solar and a wide range of IMFs.
    We also combine the synthetic SEDs with the CLOUDY photoionization code for more accurate predictions of nebular emission,
    and to study possible departures from case~B assumed in the synthesis models.}
   {The ionizing fluxes, UV continuum properties, and predicted \lya\ and \Heiiuv\ line strengths are 
    presented for synthesis models covering a wider range of parameter space than our earlier calculations.
    Strong departures from case~B predictions are obtained for \lya\ and 2$\gamma$ continuum at low metallicities.
    At low nebular densities both are shown to be enhanced proportionally to the mean energy carried by 
    the Lyman continuum photons emitted by the ionizing source. Larger \lya\ equivalent widths are therefore
    predicted at low metallicity.
    The \Heiiuv\ line can be weaker than case~B predicts (in terms of flux as well as the equivalent width) 
    due to its ionization parameter dependence and to the enhanced underlying 2$\gamma$ continuum.}
   {Our results have implications for the interpretation of star-forming metal-poor and/or high redshift galaxies,
    for galaxies among the \lya\ emitters (LAE) and Lyman Break galaxy (LBG) populations, and for searches
    of Population III stars in the distant Universe.}


   \maketitle
   %
\section{Introduction}
\label{s_intro}
Over the last decade, the execution of deep multi-band imaging surveys like GOODS
\citep{Vanzella05,Vanzella06,Vanzella08,Vanzella09,Popesso09, Balestra10} has resulted in catalogues of significant numbers of
galaxies with photometric redshifts greater than five when the
Universe was only a little over a Gyr old. In some cases, these
redshifts have been spectroscopically confirmed by the detection of
\lya\ in emission and/or the presence of an identifiable Lyman
break. The availability of an increasing sensitivity in the NIR,
notably with the newly-installed WFC3 camera in the HST and with 
the JWST in the near future, opens the
possibility of selecting high quality candidates up to and beyond a
redshift of 10. This is the epoch where it can be expected that
stellar populations have a very low metallicity, which may result
in an excess of hot, high mass stars radiating strongly in the Lyman continuum.

To provide appropriate spectral templates for such metal-poor
star-forming galaxies and to predict the observable properties of
starbursts with primordial and more evolved chemical compositions
\citet{Scha02,Scha03} has computed new evolutionary synthesis models
and has demonstrated the importance of nebular emission (lines and
continua) at low metallicity. 
These and other studies \citep[see e.g.][]{Tumlinson00,Tumlinson01,Bromm01} 
have in particular highlighted the use of strong \lya\ emission and nebular 
\heii\ emission to search for objects containing Population~III (hereafter Pop~III)
stars. Since then, various searches for the \Heiiuv\ signature from Pop~III have
been undertaken at different redshifts, yielding so far non-detections and 
interesting upper limits on the Pop~III star-formation rate density 
\citep[see][ and references therein]{Schaerer08,Nagao08}.
Furthermore, among the numerous surveys for \lya\ emitters
at different redshifts, some studies have found objects with 
apparently unusually strong \lya\ emission (high equivalent widths),
which could be indicative of very metal-poor (even Pop~III) stellar populations
or unusual IMFs \citep[see e.g.][]{Malhotra02,Yamada05}.
Other groups have invoked ``unusual'' IMFs, extremely metal-poor stellar populations,
and/or leakage of Lyman continuum radiation to explain the apparently very blue UV slopes 
found for some very high redshift ($z \sim 7$) galaxies \citep{Bouwens10_betaz7}. 
However, the significance of these results is questionable, and the present
data does not require such  ``non-standard'' assumptions \citep{SdB10,Fin10}.
In any case, it is of interest to examine how reliable some of the major observables
predicted by standard evolutionary synthesis are.

Indeed, a shortcoming of evolutionary synthesis models such as the ones mentioned above
is that they calculate nebular emission in an approximate manner assuming simplified
physics, such as case~B recombination theory \citep[cf.][]{Osterbrock06},
and constant emissivities for adopted constant values of the 
electron temperature and density in the \hii\ region surrounding
the starburst. In fact, as demonstrated in this paper, significant departures 
from case B are expected at low metallicities leading to stronger \lya\
emission, and the strength of nebular \heii\ emission predicted
by full photoionization models can be reduced with respect to simple
recombination theory. 
Indeed these physical effects, related to an increased importance of 
collisional effects at low metallicity due to lower radiative cooling 
and harder ionizing spectra -- for \lya\ -- and due to competition
between H and He for ionizing photons -- for the intensity of \heii/H -- 
have been known for a while in studies of metal-poor \hii\ regions 
\citep[cf.][]{Davidson85,Stasinska99, Luridiana03} and planetary nebulae 
\citep{Stasinska86}. \citet{Panagia02,Panagia05} has recently explored 
photoionization models for primordial nebulae.
However, the importance of the above effects for the UV emission
lines has so far not been thoroughly examined,
in particular in the context of metal-poor and distant starburst galaxies
and using up-to-date evolutionary synthesis models.
The photoionization models presented here, combined with
our evolutionary synthesis models, are intended to provide a framework 
within which to improve our knowledge of primeval
galaxies and related objects.

Another limitation of the synthesis models of \citet{Scha02,Scha03} 
concerns the initial mass function (IMF). For simplicity, three
different choices of the IMF were adopted for the bulk of the calculations
in these papers. However, different IMFs have been suggested
in other studies related to Pop~III and early stellar generations,
and considerable uncertainties remain on the true shape of the 
IMF in the early Universe and its dependence (or not) on 
physical parameters.
To enable the examination of the effects of a broader choice of IMFs
on the expected observable properties of starbursts, we here extend
the calculations of \citet{Scha03} to eight different IMFs.
The resulting model grids, available in electronic format, 
should provide state-of-the-art predictions for the interpretation
of high redshift galaxies, to estimate their contribution to
cosmic reionization, and for other topics.

The paper is structured as follows. In Sect.\ \ref{s_models}
we describe the input physics and the model calculations 
with our evolutionary synthesis code and with the photoionization
code CLOUDY. The predictions from the synthesis models 
concerning the UV continuum, the ionizing flux, \lya\ and \Heiiuv\ emission
are presented in Sect.\ \ref{s_uv_synthesis}. In Sect.\ \ref{s_cloudy}
we discuss the results from the photoionization models
using black body spectra, explain the deviations from case B and
provide simple formulae to describe these effects on \lya. 
In Sect.\ \ref{s_sed} we show how to connect realistic SEDs 
with results from photoionization models using black body ionizing spectra.
Our results and several implications are discussed in Sect.\ \ref{s_discuss}.
The main results are summarised in Sect.\ \ref{s_conclude}.

\section{Modeling techniques}
\label{s_models}

\subsection{Synthesis models}
\label{s_synthesis}

We have used the evolutionary synthesis code of \citet{SV98}.
with the physical ingredients (stellar tracks, atmospheres, and prescriptions
for nebular line and continuum emission) from \citet{Scha03}.
In particular these models allow us to predict the integrated properties
of stellar populations at all metallicities from zero (Population III) to
``normal'', solar-like metallicity. The computations have been
done for the same metallicities as in \citet{Scha03}.

\begin{table*}[htb]
\caption{
Summary of IMF model parameters.
Model ID is the label used in the Figures, colour code the colour and linestyle.
Note: the definition of $M_c$ and $\sigma$ is as in \citet{Tumlinson06}.
In particular sigma is the variance in $\ln(m)$, not $\log(m)$!}
\label{t_imf}
\begin{tabular}[htb]{llrrrrrll}
Model ID & colour code & \mlow\ & \mup\ & $\alpha$ & $M_c$ & $\sigma$ & reference/comment & ID in files
\\ \smallskip 
\\ \hline
Salpeter & black         & 1 & 100 & 2.35 &   &  & A in \citet{Scha03} & S \\
B        & green, dashed & 1 & 500$^a$ & 2.35 &   &  & B in \citet{Scha03} & B \\
C        & cyan, dashed  & 50& 500$^a$ & 2.35 &   &  & C in \citet{Scha03} & C \\
Scalo    & blue          & 1 & 500$^a$ & 2.7$^b$ & & & \citet{Scalo86}         & Sc \\
TA       & red           & 1 & 500$^a$ &      & 10. & 1.0 & A in \citet{Tumlinson06} & TA \\
TB       & magenta       & 1 & 500$^a$ &      & 15. & 0.3 & B in \citet{Tumlinson06} & TB \\
TE       & yellow        & 1 & 500$^a$ &      & 60. & 1.0 & E in \citet{Tumlinson06} & TR \\
L05      & blue, dashed  & 1 & 100 &      & 5.  &     & \citet{Larson98} & l0 \\
 \hline
\multicolumn{9}{l}{$^a$ For metallicities $Z\ge 0.0004 = 1/50 \zsun$, we adopt
$\mup = 120 \msun$, the maximum mass for which Geneva stellar evolution tracks 
are available.}\\
\multicolumn{9}{l}{$^b$ Power-law exponent for $M \ge 2 \msun$.}
\end{tabular}
\end{table*}

\subsubsection{Stellar initial mass function}
The main extension presented here with respect to the calculations of \citet{Scha03}
concerns different assumptions regarding the stellar IMF.
A wide range of IMFs has been considered, including power-law IMFs, 
such as the Salpeter or \citet{Scalo86} IMFs, log-normal IMFs, and the 
\citet{Larson98} IMF. The corresponding parameters are summarised
in Table \ref{t_imf}.
The stellar mass range is defined by the lower and upper mass cut-offs,
\mlow\ and \mup\ respectively. $\alpha$ being the slope of the power-law.
The log-normal IMFs are described by the characteristic mass $M_c$ and
its dispersion $\sigma$. The cases computed here correspond to 
the cases A, B, and E in the chemical evolution study of \citet{Tumlinson06}.
The \citet{Larson98} IMF is described by a single parameter, its
characteristic mass $M_c$. We have computed one such case, assuming
the same value of $M_c$ as \citet{Ciardi2001} in their reionization
calculations.
Note, that at $Z \ge 0.0004 = 1/50 \zsun$ the upper mass cut-off is set to $\mup = 100$
or $120 \msun$ for all IMFs, since tracks for more massive stars are not available.
The quantities discussed here are insensitive to assumptions on the IMF at low
masses. Our absolute quantities may therefore simply be rescaled to other
IMFs including e.g.\ an extension below 1 \msun.

Current knowledge suggests that the IMF for massive stars is close to Salpeter 
with an upper limit of $\mup \sim$ 100--120 \msun\ from solar metallicity
down at least to $\sim$ 1/50 \zsun, and that a qualitative shift of the
IMF towards higher characteristic masses occurs below a
critical metallicity of the order of $Z_{\rm crit} \sim 10^{-5\pm 1} \zsun$ 
\citep{schne02,schne03}.

\subsubsection{Star formation histories}
For each metallicity $Z$ and IMF we have computed evolutionary synthesis models
for the two limiting cases of 
1) an instantaneous burst, and 2) constant star formation (CSFR).
Results for other star formation histories can be derived from the electronic
files for the simple stellar population (burst) models (see Sect.\ \ref{s_output}).
In both cases the calculations have been carried out with a small time step
(0.1 Myr) to ensure the accuracy of the time integrated quantities for 
the CSFR case. The calculations have to be carried out up to ages
of 1 Gyr. This covers the allowed ages and of galaxies at redshifts $z \ga 5.8$,
of interest here, as well as sufficiently long star formation timescales
to reach equilibrium in various observable properties (cf.\ below).
\subsubsection{Nebular emission}
To include nebular emission (recombination lines and continuum processes) in our
synthesis models we initially make the following ``standard'' simplifying assumptions
\cite[see][]{Scha02,Scha03}:
ionization bounded nebula, constant electron temperature and density ($T_e$, $n_e$),
and case~B. Case~B in particular assumes that the recombinations to the ground-state 
immediately yield locally another ionization, and that photoionizations occur only from the
ground-state. As we will see below, the latter may not be true in very metal-poor nebulae,
leading to significant changes in the predicted spectrum of hydrogen.
With these assumptions both the H and He recombination lines
as well as nebular continuum emission (including free-free and free-bound emission
from H, neutral He, and singly ionized He, and two-photon emission of H) are fully specified
and their luminosity is proportional to the ionizing photon flux $Q$ in the appropriate
energy range.
To reflect to first order the changes of the conditions in the \hii\ regions with metallicity,
the value of the line luminosity coefficient and nebular continuous emission
coefficients at metallicities $Z/\zsun < 10^{-3}$ are changed as in \citet{Scha03}.
More precisely we adopt $T_e=$ 30 (20) kK for lines (continua) $Z/\zsun < 10^{-3}$ and $T_e$=10 kK
for higher metallicities, and a low ISM density $n_e=100$ cm$^{-3}$.

For the \lya\ luminosity we have, with the assumptions just spelled out:
\begin{equation}
\label{eq_lyacaseb}
  L_B(\lya) = (1 - f_{\rm esc}) \Qh \, \times h \nu_{\lya} \times \frac{\alpha^{\rm eff}_{2 ^2P}}{\alpha_B}
\end{equation}
where the index `B' stands for case~B, $\alpha_B = \alpha^{\rm eff}_{2 ^2P} +\alpha^{\rm eff}_{2 ^2S}$
is the total case B recombination coefficient, and $f_{\rm esc}$ is the escape fraction 
of ionizing photons out of the \hii\ region (or galaxy). In all Figs.\ shown in this paper
we assume $f_{\rm esc}=0$.
In typical conditions $\alpha^{\rm eff}_{2 ^2P} / \alpha_B \approx$ 0.6--0.7. In other words
approximately 2/3 of the Lyman continuum photons give rise to the emission of a \lya\
photon, the assumption commonly made.
Similar relations also hold for other recombination lines, such as \Heiiuv, which
is of special interest here.
In our ``standard'' synthesis models we simply assume
\begin{eqnarray}
\label{eq_lya}
  L_B(\lya)   & = & Q({\rm H})  \, \times c_1,  \\
\label{eq_heii}
  L_B({\rm He~{\sc II} \lambda 1640}) & = & Q({\rm He^+})  \, \times c_2, 
\end{eqnarray}
with $c_1= 1.04 \times 10^{-11}$ erg, and  $c_2= 5.67 \times 10^{-12}$ ($6.04 \times 10^{-12}$) erg
for $Z \le 1/50$ \zsun\ ($>1/50$ \zsun), and for $f_{\rm esc}=0$. The atomic data is from 
Hummer \& Storey (1995) for low densities \citep[cf.][]{Scha03}, and \Qh\  and \Qhep\ are
the ionizing photon flux (in photon s$^{-1}$) above 13.6  and 54 eV respectively.

Continuous nebular emission including free-free and bound-free emission by H,
neutral He, He$^+$, and He$^{+2}$, as well as the two-photon continuum
of hydrogen is included as described in \citet{Scha02}, assuming $T_e=20$ kK for $Z/\zsun \le 5.\times 10^{-4}$
and $T_e=10$ kK otherwise.

As we will show below (Sect.\ \ref{s_cloudy}), a proper treatment 
of all relevant processes leads to significant deviations from case B at 
very low metallicities, increasing in particular the \lya\ luminosity, $L(\lya)$. 
In this case the $L(\lya)$ can be rewritten as
\begin{equation}
L(\lya) = L_B(\lya)\times  P \times \frac{\tilde{f}_{\rm coll}}{2/3},
\label{eq_caseb_corr}
\end{equation}
where $P$ and $\tilde{f}_{\rm coll}$ are terms describing the mean photon energy
in the Lyman continuum, and accounting for for collisional effects at high density.
To compute the \lya\ equivalent width $W(\lya)$ we proceed as in 
\citet{Scha02,Scha03}, where we use the continuum
flux at 1215.67 \AA\ obtained from linear interpolation of the total
(stellar + nebular) continuum (in $\log$) between 1190 and 1240~\AA,
chosen to avoid underlying stellar \lya\ absorption and other absorption
lines if present. The predicted stellar \lya\ absorption is small compared 
to the emission, except for ``post-starburst'' phases \citep[see e.g.\ Fig3.\ in ][]{SV07}.

\begin{figure*}[htb]
\centerline{\psfig{figure=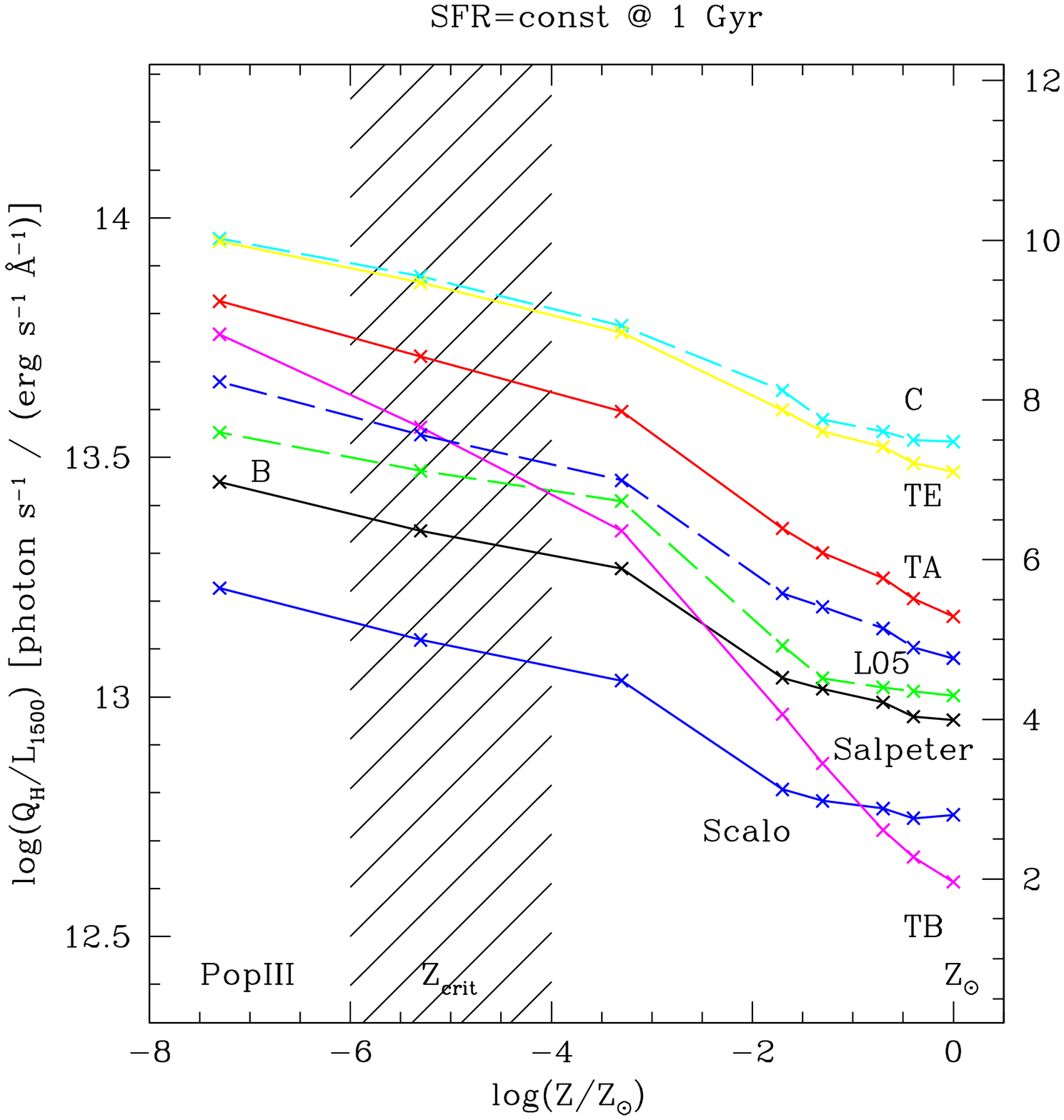,width=8.8cm}
	    \psfig{figure=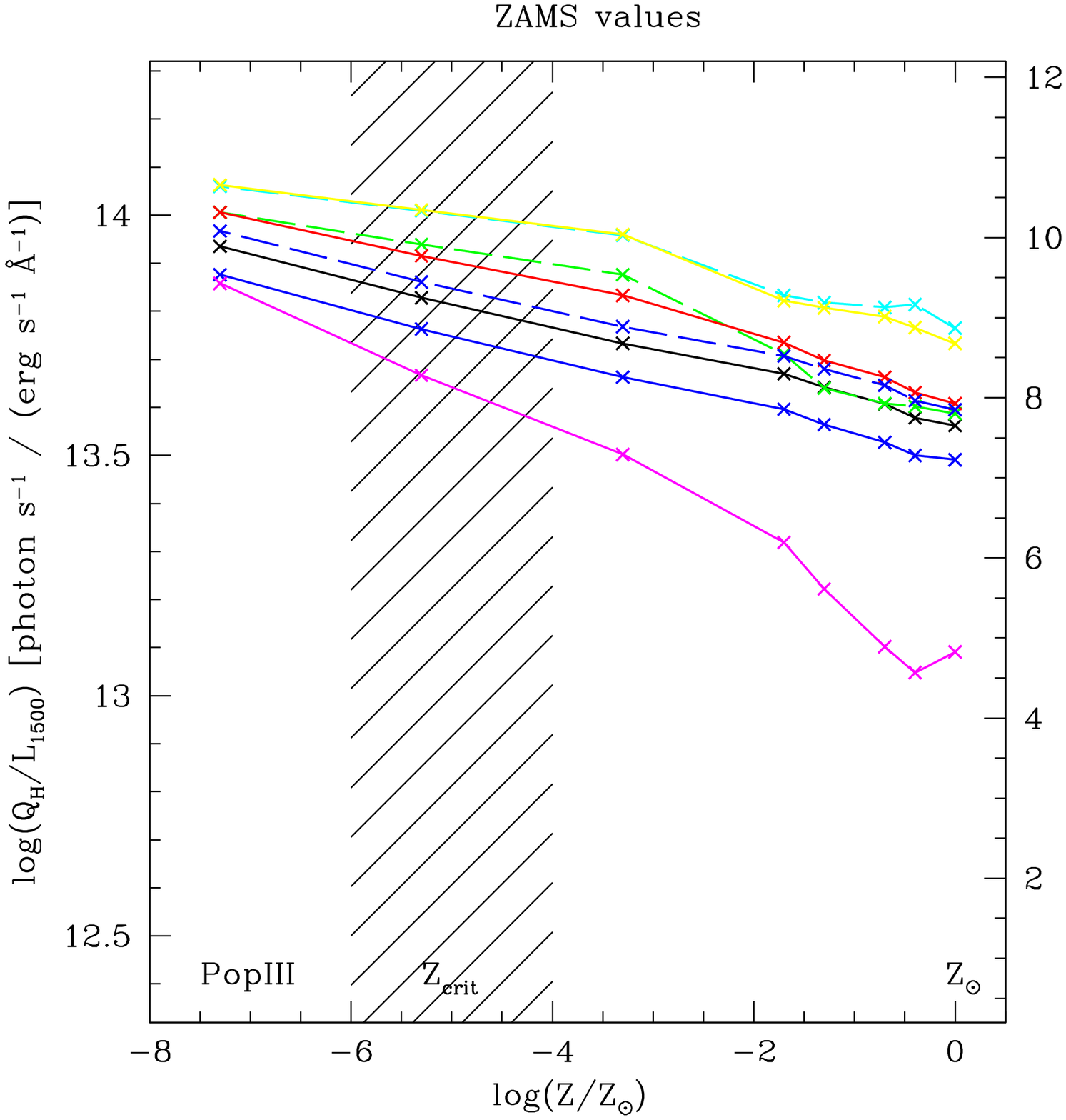,width=8.8cm}}
\caption{Relative output of hydrogen ionizing photons to UV continuum light, measured
at 1500 \AA\ restframe, $Q_H/L_{1500}$, as a function of metallicity 
for constant star formation over 1 Gyr (left panel) and very young 
bursts (right panel).
$Q_H/L_{1500}$ is given in $L_\lambda$ units on the left side of each
panel, and in $L_\nu$ units on the right.
Results for different IMFs are shown using the colour codes and labels
summarised in Table \ref{t_imf}.
The shaded area indicates the critical metallicity range where
the IMF is expected to change from a ``normal'' Salpeter-like regime to   
a more massive IMF (see text).
}
\label{f_nlyc}
\end{figure*}

\subsubsection{Model output}
\label{s_output}
Our evolutionary synthesis code predicts a large variety of observable and related quantities derived 
from the detailed synthetic spectra. Here we focus on mostly on quantities describing the 
spectrum in the Lyman continuum, the UV (rest-frame) spectrum, as well as the \lya\
and \Heiiuv\ emission lines. The full set of model results, including also numerous other quantities 
not discussed in this paper, are available in electronic format upon request to one of the authors (DS), 
on {\tt http://obswww.unige.ch/sfr}, or via the CDS.

\subsection{Photoionization models}
\label{s_photo}

To predict more accurately the nebular emission from starbursts and to investigate possible departures 
from the simplified assumptions made in our synthesis models we use the photoionization code CLOUDY version 08  \citep{Ferland98}.
The models we consider are ionization-bounded with a closed, spherical geometry and constant density.
These assumptions imply in particular that all ionizing photons are absorbed in the \hii\ region, i.e.\
$f_{\rm esc}=0$. In certain circumstances, especially in high redshift galaxies, a fraction of
the Lyman continuum photons are expected to escape 
\citep[see e.g.][]{Gnedin08,Wise09,Razoumov09}. To first order the results obtained in this paper
can simply be rescaled to such cases, as discussed below.
The main input parameters of our models are: 
the spectral energy distribution (SED) of the ionizing source, the nebular density (given by
$n_H$), the hydrogen number density), the ionization parameter ($U$), and the nebular metallicity (Z$_{\rm neb}$).
For the SED we adopt black-body spectra described by T$_{bb}$ and SEDs from our evolutionary synthesis models.
The ionization parameter (at the inner edge of the cloud) is defined as:
\begin{equation}
U = \frac{Q(H)}{4 \pi r_{\rm in}^2 \times n(\rm H) \times c}
\end{equation}
where $r_{\rm in}$ is the inner radius of the nebula which has been kept constant in our models (10$^{17}$ cm) and $c$ is the speed of light.
The small inner radius used in our calculations results in a sphere-like (not shell-like) geometry of the nebulae.
The ionization parameter can change throughout the nebula (decrease towards outer parts).

First, we have calculated the grid of photoionization models using the photoionization code CLOUDY covering
$\log(U)$~=~-4, -3, -2, -1; T$_{bb}$~=~40,000--150,000 K; $\log(n(\rm H))$~=~1, 2, 3, 4  \cmc\ for a 
primordial nebula and a number of higher metallicities. In total 192 models were computed for each metallicity.
In order to keep the same ionization parameter while changing the density of the gas, we adapt the number of ionizing photons (keeping
the shape of the SED).
The metallicity is defined by scaling the solar abundance pattern.

Note that the electron temperature is neither constant nor fixed in our models. 
Its spatial distribution results from the computation of each
photoionization model and it is a function of depth in the nebula.
In practice it depends on all the parameters that are being investigated (T$_{bb}$, $n_H$, $U$, Z$_{neb}$).
For the coolest primordial model (the coolest black body, the lowest density and the lowest ionization parameter) the temperature 
in the inner part of the cloud is around 12,000 K and for the hottest ones it reaches $\sim$ 38,000 K.

Selected models were subsequently computed using the SEDs from the synthesis models described above.

\section{Predicted UV properties from synthesis models}
\label{s_uv_synthesis}
We now present and discuss one-by-one the main predictions of our synthesis models
for different IMFs (see Table \ref{t_imf}), for metallicities from zero (Pop~III) to
solar, and for two different limiting star-formation histories (bursts and SFR=const).
Since properties of stars below $Z \la 10^{-9}$ (i.e.\ $Z/\zsun \la 10^{-7.3}$)  essentially
converge to those of metal-free stars we assign this metallicity value to Pop~III stars,
as in \citet{Scha03}.

Note that all UV continuum predictions from the synthesis models
described in this Section are based on the simplified assumptions
spelled out above to compute nebular emission. This implies in particular
that at low metallicities the contribution from the two-photon continuum
process should be higher, increasing thus e.g.\ the predicted 
UV luminosity at 1500 \AA, for the reasons discussed in Sect.\ 
\ref{s_cloudy}.

\subsection{Ionizing photon production}
A quantity of interest, e.g.\ to determine the contribution of 
galaxies to cosmic reionization, is the 
the relative output of hydrogen ionizing photons to observable UV light.
Here we provide $Q_H/L_{1500}$, where the Lyman continuum
flux $Q_H$ is expressed in units of photon s$^{-1}$, and the UV restframe
luminosity at 1500 \AA\ is $L_{1500}$ in $L_\lambda$ (\ergs\ $\AA^{-1}$)
or  $L_\nu$ (\ergs\ Hz$^{-1}$) units
\footnote{The transformation is $\log(Q_H/L_\lambda( 1500 \AA)) =
\log(Q_H/L_\nu(1500 \AA))- 12.12$.}.
Alternatively, to derive the Lyman continuum output per unit stellar
mass (or per baryon),  $Q_H/L_{1500}$ discussed here can be combined
with the ratio SFR$/L_{1500}$ given below, or can directly be derived
from the available data files.

\begin{figure}[ht!]
\centerline{\psfig{figure=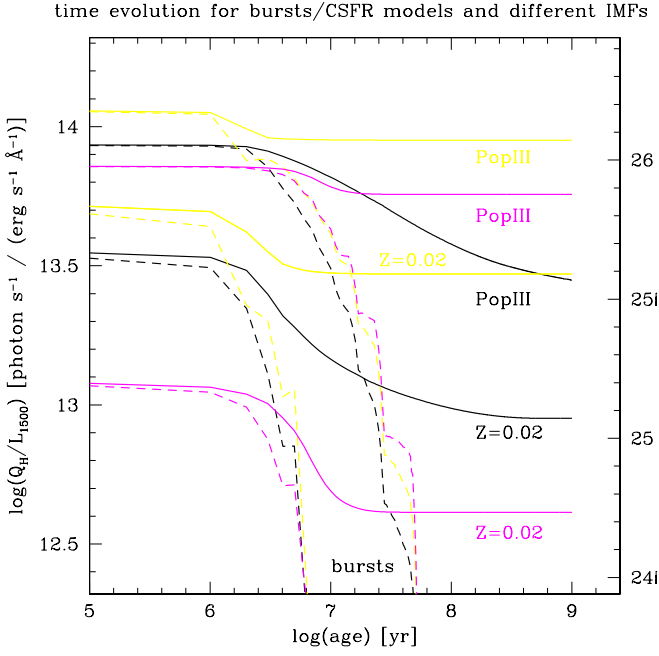,width=8.8cm}}
\caption{Temporal evolution of $Q_H/L_{1500}$ for selected IMFs (Salpeter, TB, TE,
colour-coded as in previous Figs.)  and metallicities (Pop~III, \zsun, labeled).
The solid curves show the time evolution for constant SFR models
towards their equilibrium value, the dashed curves instantaneous
burst models. See text for discussion.}
\label{f_time}
\end{figure}

\begin{figure}[ht!]
\centerline{\psfig{figure=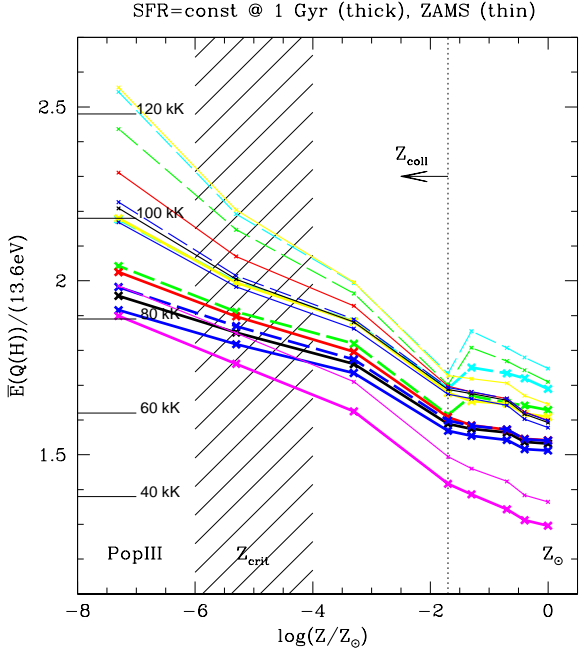,width=8.8cm}}
\caption{Mean ionizing photon energy in units of 13.6 eV as a function
of metallicity, shown for SFR=const (thick lines) and for the ZAMS (thin lines).
Results for different IMFs are shown using the same colour codes as in
Fig.\ \ref{f_nlyc} (cf.\ Table \ref{t_imf}).
Lined and labels on the left indicate the corresponding blackbody 
temperatures. 
\zcoll\ shows the approximate metallicity limit below with collisional effects 
lead to significant departures from case B.}
\label{f_eq}
\end{figure}

\begin{figure}[hb!]
\centerline{\psfig{figure=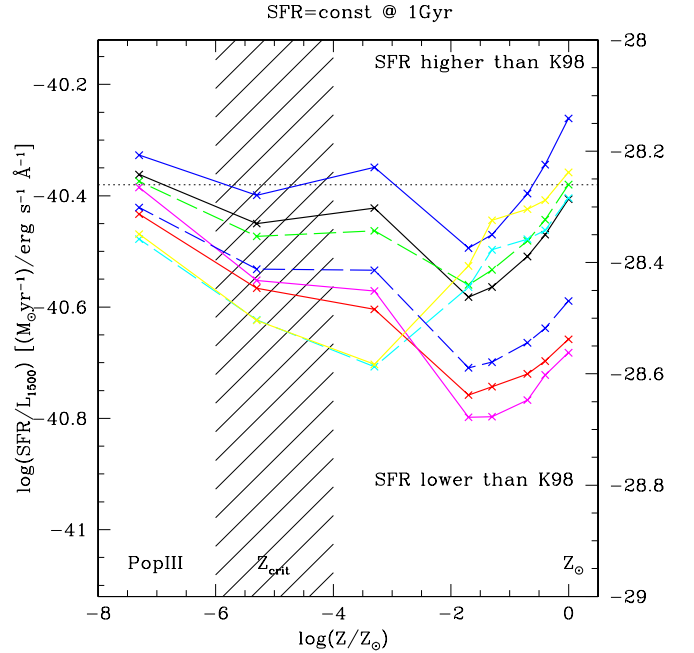,width=8.8cm}}
\caption{Dependence of the SFR(UV) calibration on metallicity and IMF.
Shown is the SFR per unit UV luminosity at 1500 \AA\ in 
units of \msunyr\ per (\ergs\ $\AA^{-1}$) on the left y-axis, or
in \msunyr\ per (\ergs\ Hz$^{-1}$) on the right y-axis.
Same symbols and colours as in previous Figures (cf.\ Table \ref{t_imf}).
These values assume constant SF over 1 Gyr.
The dotted horizontal line show the value of the Kennicutt (1998)
SFR(UV) calibration rescaled to a Salpeter IMF with $\mlow = 1 \msun$
for comparison. Above/below this line the SFR deduced from the UV luminosity
is higher/lower.}
\label{f_sfr}
\end{figure}

In Fig.\ \ref{f_nlyc} we show $Q_H/L_{1500}$
as a function of metallicity for constant star formation over 1 Gyr
(CSFR, left panel), and very young ($\la$ 0--4 Myr) populations (right panel).
As expected $Q_H/L_{1500}$ increases with decreasing metallicity,
since the ionizing flux depends very strongly on the effective
stellar temperature and hence increases more rapidly than the UV
luminosity. The IMF dependence also behaves as expected, with the 
IMFs favouring the most massive stars showing also the highest
the $Q_H/L_{1500}$ ratios, since $Q_H$ increases more rapidly with
stellar mass than the UV luminosity.
Notable is actually the increase of  $Q_H/L_{1500}$ from \zsun\
to $\sim 10^{-4} \zsun$, where no major change of the IMF is expected
(and a Salpeter IMF is favoured).
For CSFR and for a fixed IMF, the increase of the relative ionizing
power from solar metallicity to Pop~III is typically $\sim$ 0.4--0.5 dex, 
i.e.\ a factor 2 to 3.
When considering an IMF change from Salpeter to a massive IMF
(i.e.\ all cases except Salpeter and Scalo) the increase of  $Q_H/L_{1500}$ 
is larger, approximately 0.6 to 1 dex between solar and zero metallicity.
Only for the ``TB'' IMF, a narrow, log-normal mass function peaked 
at $M =15 \msun$, we find a more extreme dependence on metallicity.
This is precisely due to the fact that this IMF singles out a  
narrow mass range, instead of averaging the metallicity dependence 
of stellar properties over a larger interval in mass.

The right panel of Fig.\ \ref{f_nlyc} shows a narrower
range of $Q_H/L_{1500}$ for zero age or very young ($\la$ 0--4 Myr) populations.
This is natural, since in this case no ``average'' is made
over populations of very different stellar ages and hence 
over strong variations of stellar parameters.
More important is the fact that {\em higher values of $Q_H/L_{1500}$
are obtained for young populations}. This is mostly due to the fact
that such populations emit a lower UV luminosity per unit SFR
since a longer timescale is needed to reach the ``equilibrium
value'' of the UV output (cf.\ below).
The ZAMS values shown here correspond to the maximum of  $Q_H/L_{1500}$
expected for stellar populations of different ages and SF histories.

To illustrate this dependence on the SF timescale, $Q_H/L_{1500}(t)$
is shown in Fig.\ \ref{f_time} for selected IMFs and metallicities.
These curves show the smooth transition from the predicted ``ZAMS''
to the CSFR values over timescales from $\sim 10^7$ yr for massive IMFs
(e.g.\ TE, TB) to $\sim$ 0.4--1 Gyr for the Salpeter IMF.
Note also that the timescale for UV properties to reach equilibrium
increases with decreasing metallicity, due to the higher
effective temperatures on the ZAMS at low $Z$.
In short, we caution that the relative ionizing photon to UV 
ratio $Q_H/L_{1500}$ may be uncertain by a factor of $\sim$ 4
depending on the SF timescale (for constant SF), or even more
for bursts.

Finally, it should be noted that 
at low metallicity the contribution of the two-photon continuum
may be larger, as shown later in Sect.\ \ref{s_cloudy}, leading to
somewhat lower values of $Q_H/L_{1500}$.
For constant SF this amounts to a decrease of $Q_H/L_{1500}$
by $\sim 0.2$ dex for Pop~III and the most extreme IMFs (TE, C),
and smaller changes otherwise.
For zero metallicity populations on the ZAMS $Q_H/L_{1500}$ should 
be reduced by $\sim$ 0.15--0.3 dex for all IMFs, and less at
higher metallicity.

\subsection{Properties of the ionizing spectra}
The properties of the ionizing spectra, such as their hardness, detailed
shape and others have already been discussed in \citet{Scha03} and shall not
be repeated here. For example, the hardness \Qrathep, measured by the 
ratio of He$^+$ ($> 54$ eV) to hydrogen ionizing ($>13.6$ eV) photons,
we predict from our new models are already bracketed by the 
the values predicted in \citet{Scha03} (see their Fig.\ 5) for the Salpeter 
and the ``C'' IMF. 
  
An interesting quantity describing the ionizing spectra is the average
energy of the photons emitted in the Lyman continuum $\overline E(Q(\rm H))$
(see definition in Eq.\ \ref{eq_emean}). This quantity and its
dependence on metallicity and IMF is plotted in Fig.\ \ref{f_eq} for
constant SF (thick lines) and for the ZAMS (thin lines).
The corresponding blackbody temperatures \tbb\ with the same
mean ionizing photon energy are also shown for illustration. 
Typically $\overline E(Q(\rm H))$ is found to $\sim$ 1.5--2.5 times
13.6 eV, the ionizing potential of neutral hydrogen, and its behaviour
with IMF, metallicity, and age behaves as expected.
These values correspond to a range of blackbody temperatures from
$\sim$ 50 to 120 kK for the hardest spectra.

On this Figure we also indicate the approximate metallicity limit
\zcoll, below which collisional effects lead to significant departures
from case B, as shown below.
In this metallicity range $\overline E(Q(\rm H))$ can also be used
to compute more accurately the intrinsic \lya\ emission line strength
(see Eq.\ \ref{eq_lya_final}).

\subsection{SFR calibrations from the UV continuum}
Figure~\ref{f_sfr} illustrates the variation of the UV luminosity
for CSFR as a function of metallicity and for the different IMFs.
Plotted is the conversion factor $c$, defined by SFR$=c \times 
L_\nu$, where $L_\nu$ is the UV luminosity in units of
\ergs\ Hz$^{-1}$, and SFR is the star formation rate in
\msunyr. As expected our model with Salpeter IMF agrees well with the 
widely used calibration from Kennicutt (1998) at \zsun\
after rescaling the latter by a factor 2.55 to account for our 
adopted value for the lower mass cut-off ($\mlow = 1 \msun$)\footnote{More explicitely the Fig.\ shows $\log c$, where 
$\log {\rm SFR}= \log c + \log L_{1500} + \log c_M$, and $c_M=2.55$
for the IMF adopted by Kennicutt ($\mlow=0.1$ \msun), or $c_M=1$ for 
$\mlow=1$ \msun.}.
This Fig.\ clearly shows that in most cases the use of the Kennicutt
calibration at low metallicity may overestimate the SFR, given
the higher intrinsic UV output of such stellar populations.
However, this may not be realistic since it relies on the assumption
of CSFR over a long timescale ($\ga 10^{8.3 \ldots 9}$ yr).
In younger populations the UV luminosity per unit SFR is lower,
and hence $c$ and the determined SFR higher (e.g.\ Schaerer 2000).

In fact, the non-monotonous behaviour of SFR$/L_{1500}$ with
metallicity observed in Fig.\ \ref{f_sfr} is due to the dependence of
the stellar contribution to the total UV output at this
wavelength. Indeed, the stellar UV output increases with decreasing
$Z$ down to $\sim 1/50$ Z/\zsun, due to the decrease of the average
temperature of stars over their lifetime.  At even lower
metallicities, however, their UV output (per unit SFR) {\em decreases}
since the bulk of their flux is emitted at $\ll 1500$ \AA
\citep[cf.\ Fig.\ 2 in][]{Scha03}. 
This implies, for a fixed IMF, a re-increase of SFR$/L_{1500}$ at $Z/\zsun \la 1/50$, 
which is only somewhat moderated by the increasing nebular contribution.
Indeed, the latter contributes typically $\sim$ 10--40\% of the flux
at 1500 \AA\ at SF equilibrium (see Fig.\ \ref{f_neb}) 
In other words neglecting the nebular continuum would lead to
differences of $\sim$ 0.05--0.15 dex in the SFR calibrations.
At low metallicity the contribution of the two-photon continuum
may be larger, as shown in Sect.\ \ref{s_cloudy}, leading to
somewhat lower values of SFR$/L_{1500}$ than shown in Fig.\ \ref{f_sfr}.
For Pop~III and the most extreme IMFs (TE, C) this implies a
downward revision of $\sim$ 0.2 dex.

\subsection{Predicted \lya\ emission}

\begin{figure*}[ht!]
\centerline{\psfig{figure=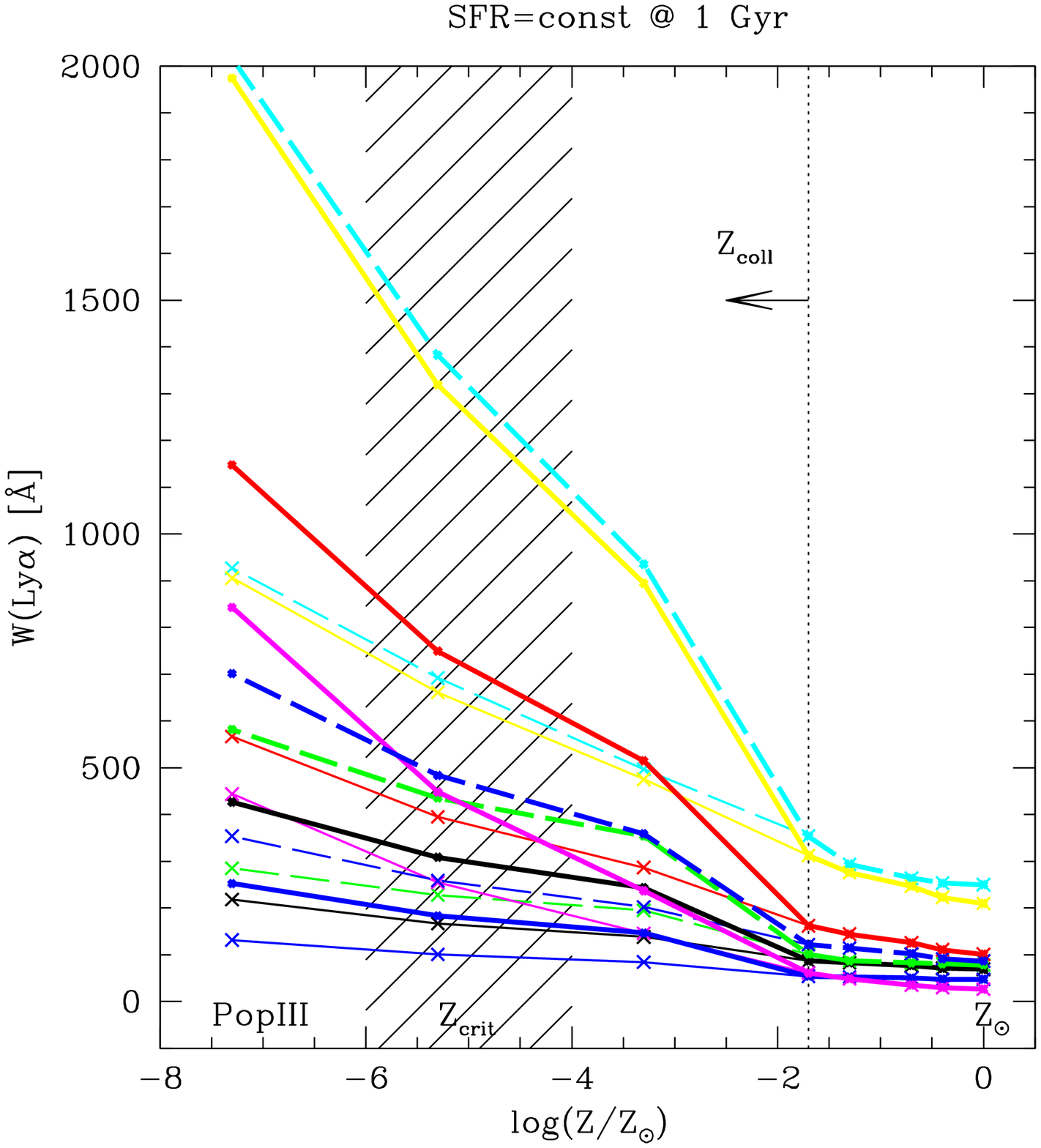,width=8.8cm}
	    \psfig{figure=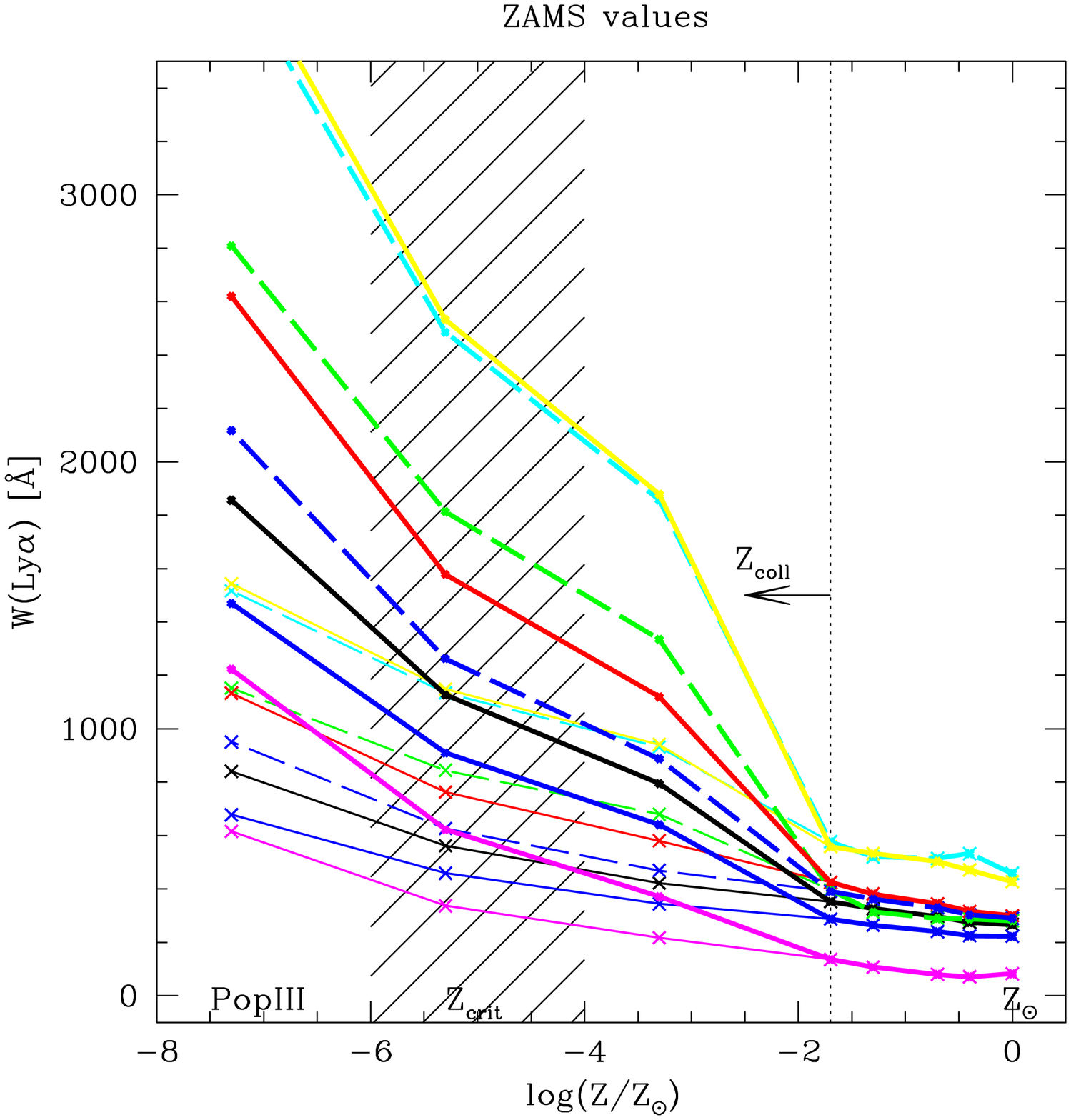,width=8.8cm}}
\caption{Predicted \lya\ equivalent width as a function of metallicity 
for constant star formation (left panel) and very young ($\le$ 1--2 Myr)
bursts (right panel). Note the different vertical scales on the two plots.
Thin lines show our ``standard'' predictions, thick
lines the predicted $W(\lya)$ accounting to first order for departure from case B
following Eq.\ \protect\ref{eq_lya_final} (assuming low density, i.e.\ 
$\tilde{f}_{\rm coll}=2/3$, and neglecting the increase of the two-photon continuum), 
leading to an increase by up to a factor $\sim$ 1.5--2.5 at low metallicities 
($Z \protect\la \zcoll$). %
Results for different IMFs are shown using the same colour codes as in
Fig.\ \ref{f_nlyc} (cf.\ Table \ref{t_imf}).
}
\label{f_wlya}
\end{figure*}

The \lya\ equivalent widths predicted by our standard models
(using Eq.\ \ref{eq_lya}) for all IMFs and metallicities 
are shown with thin lines in Fig.\ \ref{f_wlya} for constant star-formation
(left panel) and for the ZAMS (right), the latter representing
the maximum of $W(\lya)$ for each IMF.
The dependences of $W(\lya)$ are as expected from $Q_H/L_{1500}$
and the values shown here bracket those already presented in
\citet{Scha03} (with more limited variations of the IMF).
To illustrate the departure from case B found at low metallicity
from photoionization models (see Sect.\ \ref{s_cloudy}), we also show
$W(\lya)$ computed from 
Eq.\ \ref{eq_caseb_corr} in the low density
regime (i.e.\ for $ \tilde{f}_{\rm coll}=2/3$, thick lines).
Here, the equivalent widths are a factor $\sim$ 1.5--2.5 higher than our
``standard'' predictions (cf.\ Fig.\ \ref{f_eq}), since we assume 
that the continuum close to \lya\ is unchanged by this departure
from case~B.

A few words of caution about $W(\lya)$ are appropriate. First,
note that for our computations of $W(\lya)$ we use the continuum
flux at 1215.67 \AA\ obtained from linear interpolation of the total
(stellar + nebular) continuum (in $\log$) between 1190 and 1240~\AA,
chosen to avoid underlying stellar \lya\ absorption and other absorption
lines if present (cf.\ above). While $W(\lya)$ is well defined theoretically,
comparisons with observations require some caution, given the possible
complexity of the continuous spectrum shortward (due to the IGM in particular)
and longward of \lya\ (due to non-monotonic shape of the nebular continuum),
and given different choices of broadband filters (see e.g.\ the simulations
of Hayes \& \"Ostlin (2006)).

\begin{figure}[hb!]
\centerline{\psfig{figure=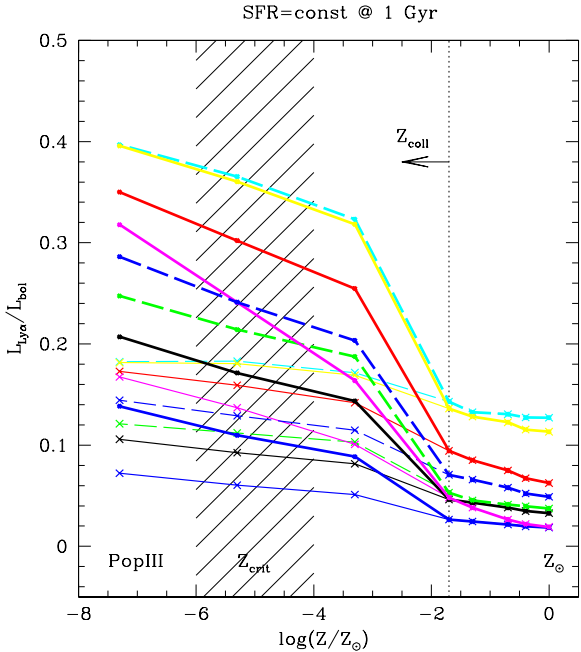,width=8.8cm}}
\caption{Fraction of the \lya\ luminosity to the total bolometric luminosity,
$L(\lya)/L_{\rm bol}$ for SFR=const as a function of metallicity and IMF. 
Results for different IMFs are shown using the same colour codes as in
Fig.\ \ref{f_nlyc} (cf.\ Table \ref{t_imf}).
Thin lines show results using our ``standard'' \lya\ predictions; thick lines the 
improved results accounting for departures from case B at very low
metallicity.  Note the resulting strong increase of our revised $L(\lya)/L_{\rm bol}$ values
from solar to very low metallicity!
}
\label{f_lyafrac}
\end{figure}

How much of the total radiative energy from the starburst is emitted
in the \lya\ line? For constant star-formation the fraction of the \lya\
to the bolometric luminosity, $L(\lya)/L_{\rm bol}$, is shown in 
Fig.\ \ref{f_lyafrac} for all metallicities and IMFs.
At solar metallicity and for a Salpeter IMF we obtain the well-known
estimate of $L(\lya)/L_{\rm bol} \sim$ 3--6 \% found in 
the first papers promoting \lya\ searches at high redshift
(e.g.\ Partridge \& Peebles 1967).
The fraction of luminosity emitted in \lya\ increases with
decreasing metallicity, due to the higher ionizing photon
flux output per unit stellar mass.
When case B departures at low metallicity are taken into account,
we find that $L(\lya)/L_{\rm bol}$ can reach up to $\sim$ 20--40\%
depending on the IMF, i.e.\ up to 10 times more than expected
from earlier calculations!
The highest values are comparable to those from the photoionization
models of \citet{Panagia02} using very hot black body spectra.
For younger populations $L(\lya)/L_{\rm bol}$ is less dependent
on the IMF than for SFR=const shown here; values of $L(\lya)/L_{\rm bol}
\sim$ 0.15--0.20 (0.35--0.40) are obtained at low $Z$ with
our standard (departure from case B) assumptions.

\subsection{\heii\ line emission from very metal-poor starburst galaxies}
\begin{figure*}[ht]
\centerline{\psfig{figure=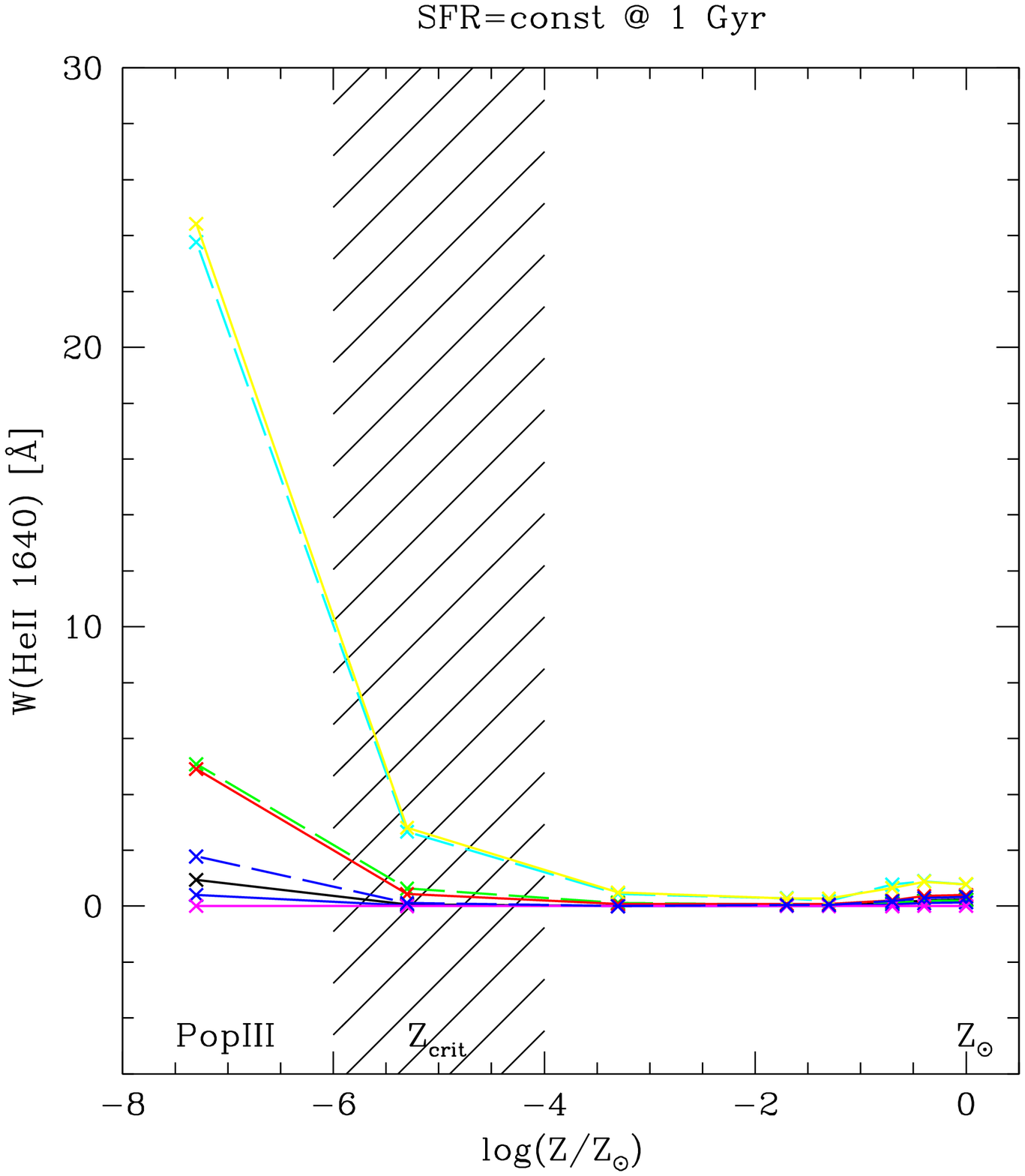,width=8.8cm}
	    \psfig{figure=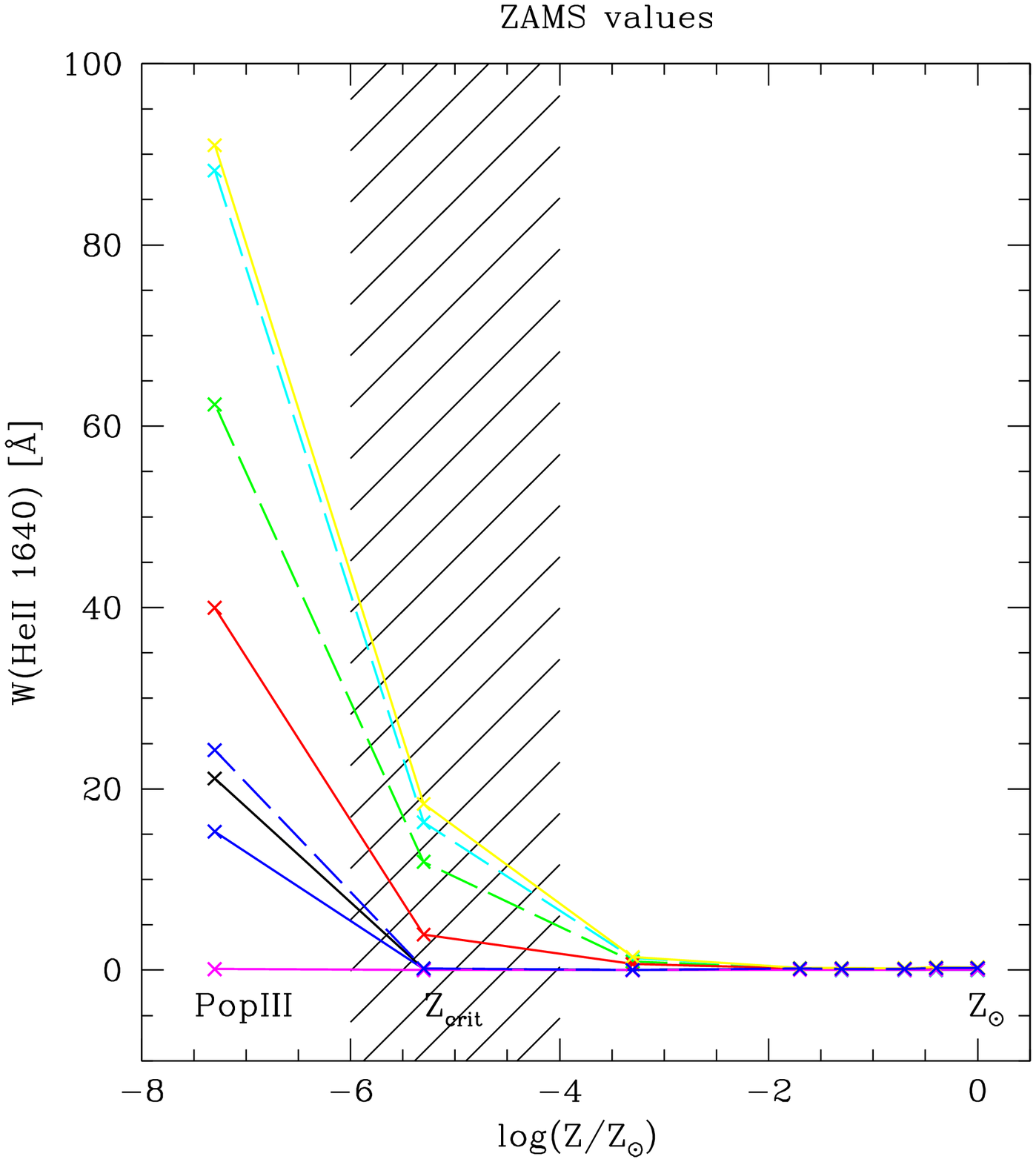,width=8.8cm}}
\caption{Predicted \Heiiuv\ equivalent width as a function of metallicity 
for constant star formation (left panel) and very young ($\le$ 1--2 Myr)
bursts (right panel). Note the different vertical scales on the two plots.
Results for different IMFs are shown using the same colour codes as in
Fig.\ \ref{f_nlyc} (cf.\ Table \ref{t_imf}). 
Note that photoionization models predict generally fainter \Heiiuv\
emission, hence lower equivalent widths, except for high ISM densities
(see Sect.\ \ref{s_heii}).
}
\label{f_wheii}
\end{figure*}

Our standard predictions for \Heiiuv\ (using Eq.\ \ref{eq_heii})
for constant star-formation and young bursts are shown in Fig.\ \ref{f_wheii}.
They complement our earlier predictions in \citet{Scha03}, and show 
the expected behaviour. Clearly, strong nebular \Heiiuv\ emission
from starbursts is only expected at very low metallicity and for
IMFs producing enough massive stars.
As we will show below, the predicted intensity of \Heiiuv\ (and 
other He$^+$ recombination lines) depends, however, also on the 
ionization parameter and on the ISM density to some extent.
Complete photoionization models predict generally fainter \Heiiuv\
emission, as discussed in Sect.\ \ref{s_cloudy}.

\begin{figure*}[htb]
\centerline{\psfig{figure=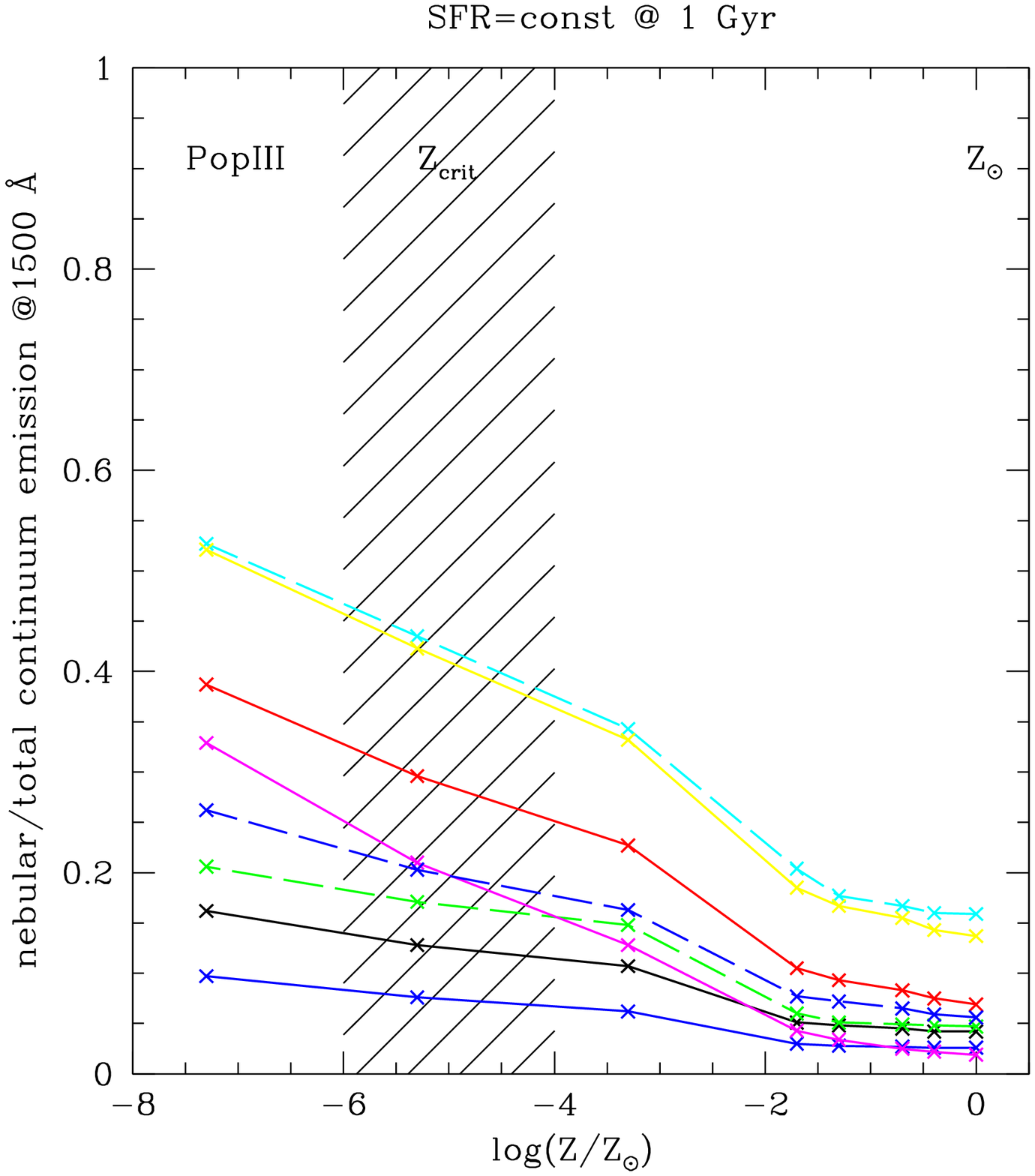,width=8.8cm}
	    \psfig{figure=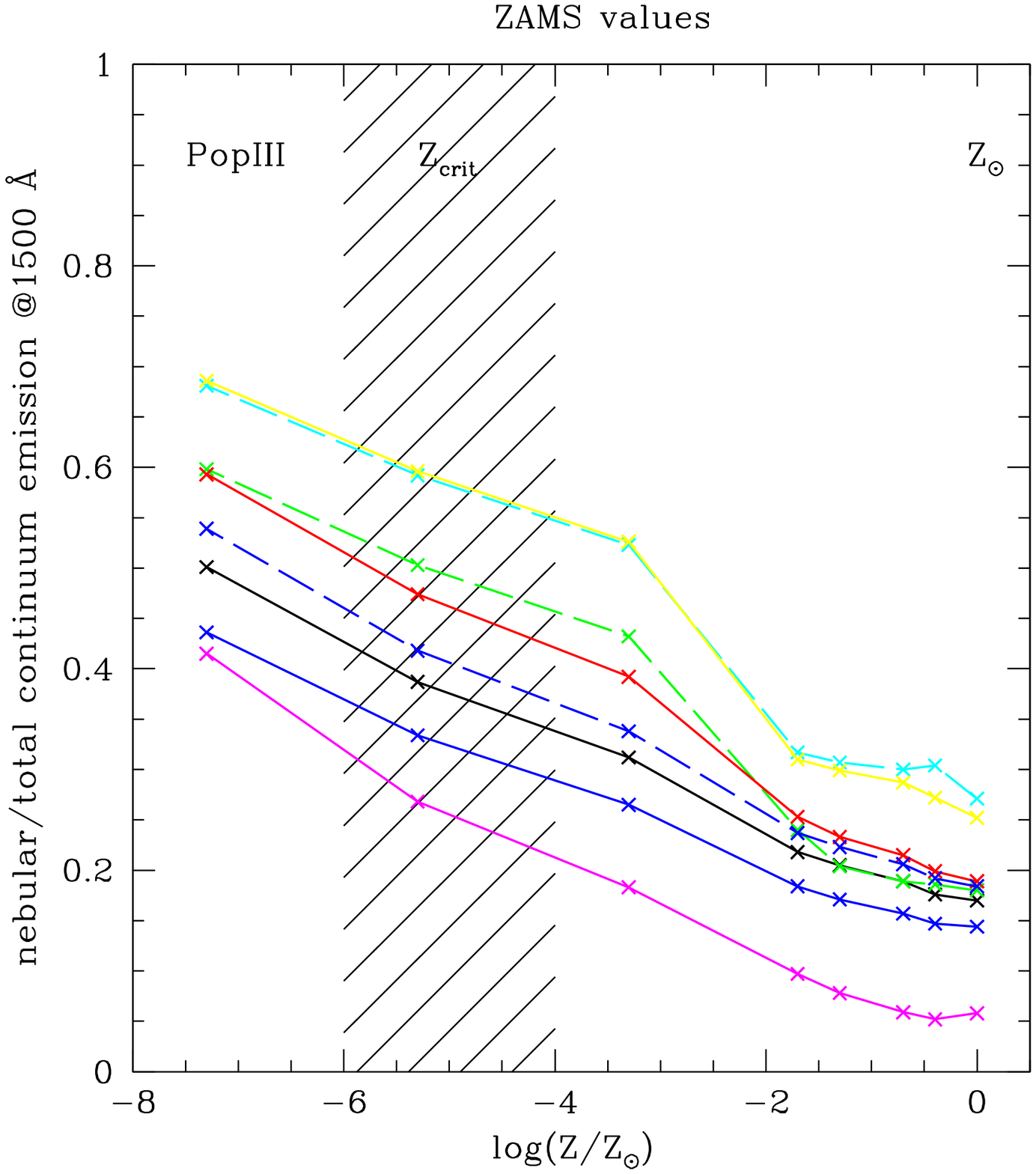,width=8.8cm}}
\caption{Contribution of nebular continuum emission to the total
emission in the restframe UV at 1500 \AA\ as derived from our 
evolutionary synthesis models.
Shown are model for constant star formation (left panel) and very young 
($\le$ 1--2 Myr) bursts (right panel).
Results for different IMFs are shown using the same colour codes as in
Fig.\ \ref{f_nlyc} (cf.\ Table \ref{t_imf}).
}
\label{f_neb}
\end{figure*}
\begin{figure*}[htb!]
\centerline{\psfig{figure=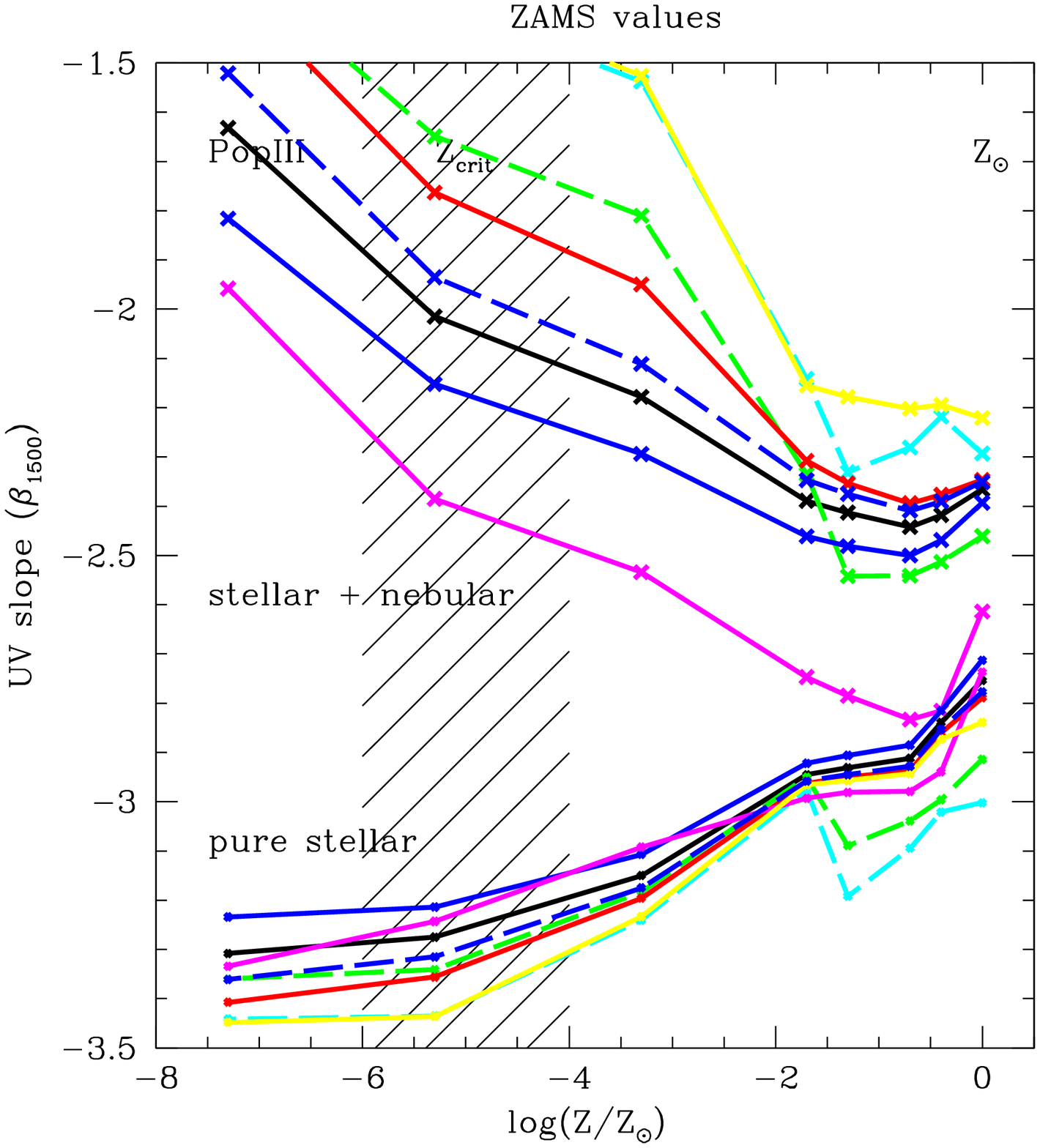,width=8.8cm}
  \psfig{figure=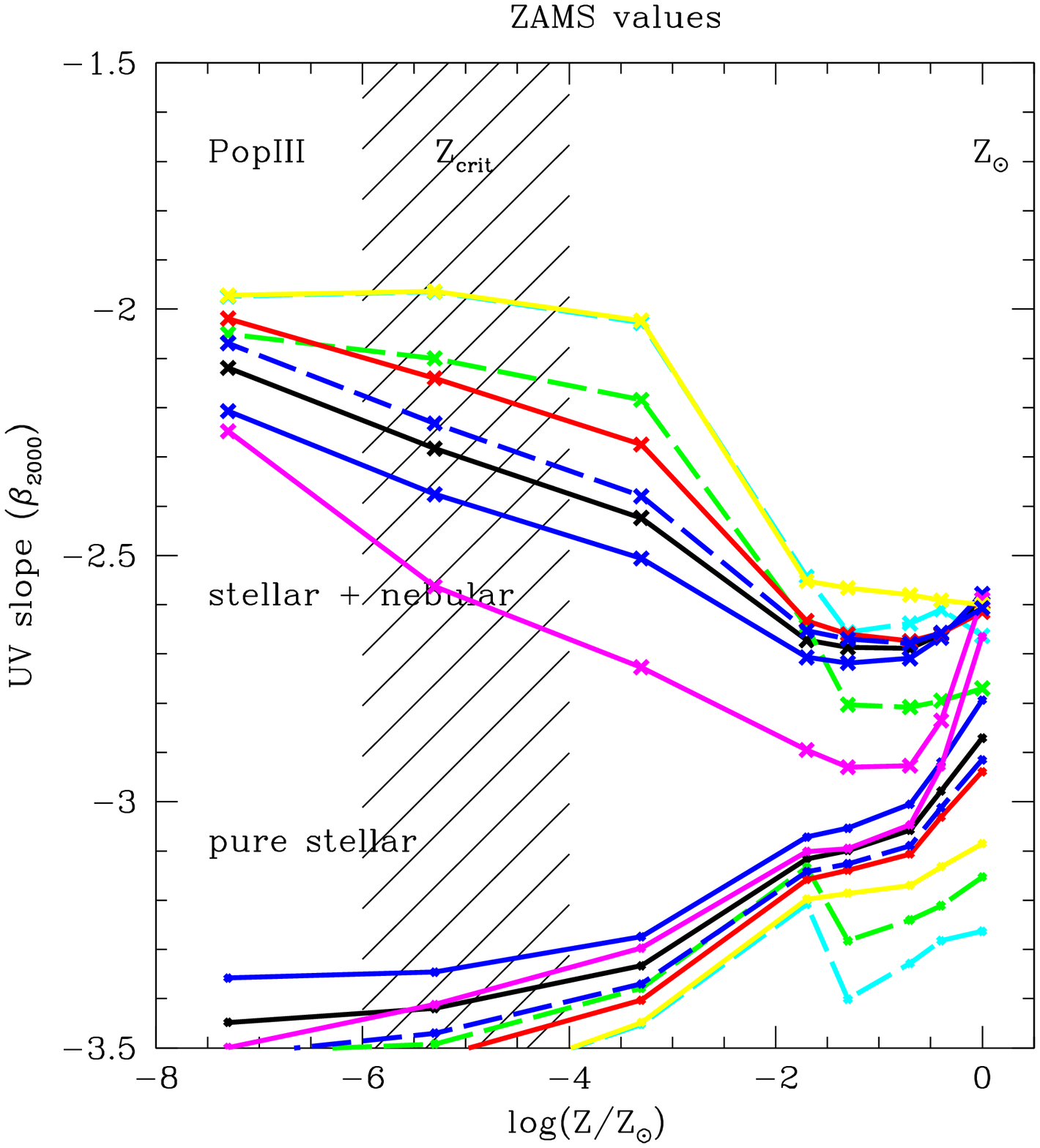,width=8.8cm}}
\caption{{\bf Left:} Predicted UV slope $\beta_{1500}$ and
for all IMFs and metallicities. The values are shown for very young (ZAMS) populations, which correspond to the bluest possible slopes (i.e.\ minimal $\beta$ values).
The upper set of lines shows the UV slopes of the total spectrum (stellar + nebular continuum),
the lower lines using the pure stellar spectrum.
Results for different IMFs are shown using the same colour codes as in
Fig.\ \ref{f_nlyc} (cf.\ Table \ref{t_imf}).
{\bf Right:} Same as the left panel, for $\beta_{2000}$. Note the difference
between $\beta_{1500}$ and $\beta_{2000}$, indicative of the deviation
of the true SED from a simple power law.
} 
\label{f_beta}
\end{figure*}

\subsection{Importance of the nebular continuum}
Figure~\ref{f_neb} shows the contribution of nebular continuous
emission to the total UV light at 1500 \AA\ for all metallicities,
IMFs, and for the usual limiting cases of star-formation 
histories.
Whereas for ``normal'' metallicities and IMFs the contribution
is relatively small ($\la$ 5\% for SFR=const), the importance 
of the nebular continuum is larger for young bursts and/or
low metallicity, as already stressed by \citet{Scha02} and \citet{Scha03}.

Since the nebular continuum at 1500 \AA\ is generally dominated by 
the two-photon continuous emission process, departures from Case~B
will lead to stronger nebular emission than shown here at very low metallicity
(see Sect.\ \ref{s_cloudy}).
To first order, the total two-photon emission is then enhanced by 
a factor $P$ for low ISM densities, increasing thus the 
contribution of the nebular continuum to the total (stellar + nebular) emission.
At high density, the two-photon emission tends to zero.
However, since the shape of the nebular continuum depends
on the detailed conditions in the nebula (density, temperature, etc.)
which are not constant as assumed in the synthesis models,
it is not possible to predict more accurately how departures from 
case~B affects the nebular continuum, without resort to
photoionization models.

\subsection{Predicted UV slope}
From our synthesis models we also measured the slope of the UV continuum
with and without nebular emission (cf.\ Schaerer \& Pell\'o 2005).
In Fig.\ \ref{f_beta} we show a condensed overview of various
$\beta$-slopes for very young populations (ZAMS), representing 
the steepest slopes, i.e.\ the minimum for $\beta$, predicted from models.
As before, the predictions are shown for all IMFs and metallicities.
Thick (thin) lines show $\beta_{1550}$ ($\beta_{2000}$), defined as the 
slope between 1300--1800 (1800--2200) \AA\ respectively\footnote{
We use the standard definition $F_{\lambda} \propto \lambda^\beta$.}.
The upper set of lines shows the $\beta$ slopes of the total 
spectrum, including stellar and nebular continuum, the lower lines
using the pure stellar spectrum. 

Clearly, the UV slope is strongly affected by nebular emission, leading to 
a significant flattening of the spectrum \citep[cf.\ Figs.\ 2 in ][]{Scha03}.
While the stellar SED becomes steeper with decreasing metallicity, the
total spectrum exhibits the opposite behaviour.
If we assume a varying contribution of the nebular continuum we may
obtain any intermediate value of $\beta$ between the ``stellar+nebular''
and ``pure stellar'' cases. This fact and the dependence of $\beta$ on
the star-formation history and age \citep[see e.g.\ Fig.\ 1 in][]{Schaerer2005},
shows that the UV slope cannot be used to determine metallicity.
For constant star-formation (not shown here), the bulk of the models show 
equilibrium values of $\beta \sim$ -2.6 to -2, quite independently of 
metallicity and IMF.

Fig.\ \ref{f_beta} also shows a difference of the order of 0.2--0.3 between
$\beta_{1500}$ and $\beta_{2000}$. Such differences may be relevant
for comparisons of the UV slope estimated from broad-band filters.
Last, but not least, the precise shape of the nebular continuum is
determined by the detailed nebular structure (i.e.\ its detailed
temperature and density structure) and can be affected by departures
from case~B, as the case of the two-photon continuum discussed in
depth below.

\section{Nebular predictions using photoionization models}
\label{s_cloudy}
\begin{figure*}[tbh]
\centering
\includegraphics[height=20cm,angle=90]{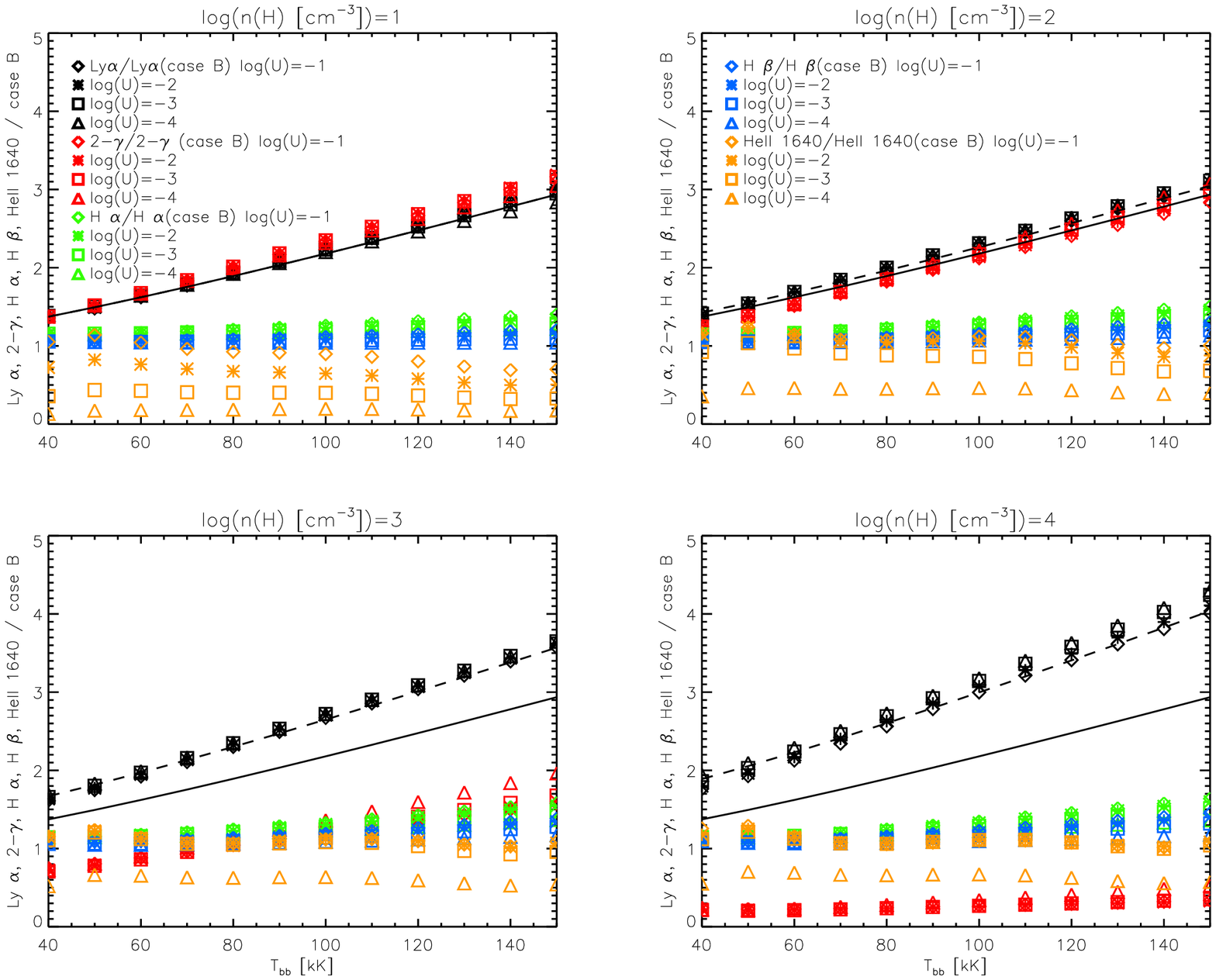}
\caption{ \lya\ (black), 2$\gamma$ (red), H${\alpha}$ (green), H${\beta}$ (blue), \heii\ 1640 \AA\ (orange)
luminosities over their predicted case~B luminosities
for the primordial nebula. The black solid line
shows $P$ (Eq.~\ref{eq_emean}), the dashed lines show $P \times \tilde{f}_{\rm coll} / \tilde{f}_{\rm coll}(\rm log(n_H=1))$
(cf.\ Eq.~\ref{eq_fcoll}).
Different symbols represent different ionization parameters, 
namely open diamonds correspond to $\log(U)=-1$, stars to $\log(U)=-2$, open squares to $\log(U)=-2$, 
and open triangles to $\log(U)=-4$.
It can be seen that \lya\ and 2$\gamma$ luminosities scale with $P$ at low densities and \lya\ is further 
enhanced at higher density. Case~B values have been 
calculated using luminosity coefficients
for $T_e$=30,000 K and $n_H$=$10^2$ cm$^{-3}$ from \protect\citet{Scha03}. For simplicity we always 
compare the results to the low density case~B limit.}
\label{all_prim}
\end{figure*}

For low metallicity nebulae ionized by very hot stars, the conventional case~B 
predictions for line and continuum emission are not good approximations to the appropriate 
nebular astrophysics. In this Section we present the predictions of detailed photoionization
modeling for such nebulae with metallicities ranging from zero (primordial = Pop~III) to solar,
explain the origin of departures from case~B,
and present a parameterisation of the results than can be readily employed for the interpretation
of low metallicity nebulae. For simplicity, we first use black body spectra as ionizing sources and discuss
later how to compare these with stellar SEDs. We have examined the nebular emission as a function of the
stellar (black body) temperature T$_{bb}$, the hydrogen number density $n_H$,
the ionization 
parameter (U), and the nebular metallicity (Z$_{neb}$), using the models described above (Sect.\ \ref{s_photo}).

Our analysis will focus mainly on the \lya\ line and the associated two-photon
continuum emission, on nebular \heii\ emission, and on the global nebular continuum.

\subsection{\lya\ line and two-photon continuum emission}
In Fig.\ \ref{all_prim} we show the deviation of the \lya\ (black symbols) and 2$\gamma$ emission
(red) and other quantities as predicted from photoionization models with primordial
composition and different densities with respect to their case~B values.
To calculate case~B luminosities we adopt Eq.\ \ref{eq_lya}
and assume $L(2\gamma)=0.5\, c_1 Q({\rm H})$ for the luminosity in the two-photon continuum. 
The numerical factor $0.5 \approx \alpha^{\rm eff}_{2 ^2S}/\alpha^{\rm eff}_{2 ^2P}$ is appropriate 
for low densities. For simplicity we always compare the results to the low density limit 
case~B predictions.

At low density, the luminosity of both \lya\ and 2$\gamma$ emission are increased 
by a factor of $\sim$ 1.3 to 3 over the black body temperature range considered here.
The physical reason for this strong departure from case~B is due to collisional effects, which
increase the population of the $n=2$ level of hydrogen from which additional
ionizations can take place, leading overall to an increased ionization rate in the nebula.
In equilibrium, this implies an increased recombination rate and
higher \lya\ and 2$\gamma$ luminosities. Collisional excitation is significant in (very)
low metallicity nebulae with hot ionizing sources because radiative cooling is reduced,
leading to higher electron temperatures. Photoionization from excited states (here from 
$n=2$), in particular, is not taken into account under case~B assumptions.

To describe quantitatively the effect of enhanced photoionization rates, including ionization from the
excited $n=2$ state, it is sufficient to consider the mean energy of the ionizing photons
in units of the ionization potential of hydrogen, $P$, given by
\begin{equation}
P = \frac{\overline E}{13.6~{\rm eV}}  =\left(\frac{ \int_{\rm 13.6~eV}^{\infty} F_{\nu}d\nu}{\int_{\rm 13.6~eV}^{\infty}\frac{F_{\nu} }{h\nu} d\nu}\right) / \left(13.6 ~{\rm eV}\right),
\label{eq_emean}
\end{equation}
where $\overline E$ is the average ionizing photon energy in the Lyman continuum in units of eV.
Indeed, as Figure~\ref{all_prim} shows, the enhancement of the
\lya\ and two-photon continuum emission, which both measure the effective
recombination rate, scales very accurately with $P$ at low density. 
This scaling shows that the available energy of the Lyman continuum 
photons -- which is in excess of the necessary minimum of 13.6 eV -- 
is ``optimally'' used to maximise the number of photoionizations,
leading to an effective increase of the ionization rate from $\propto
Q({\rm H})$ to $\propto P \times Q({\rm H})$, and hence to the same
increase in the recombination rates, i.e.\ also in the \lya\ and
2$\gamma$ luminosities.

The effect of ``boosted'' \lya\ and 2$\gamma$ emission just discussed
depends on the nebular metallicity. This is because more metals are
present and more efficient cooling results in cooler gas than in the
primordial metallicity case, decreasing the collisional effects
for H. The range over which these effects take place is discussed below.

\begin{figure}[tbh]
\centerline{\psfig{figure=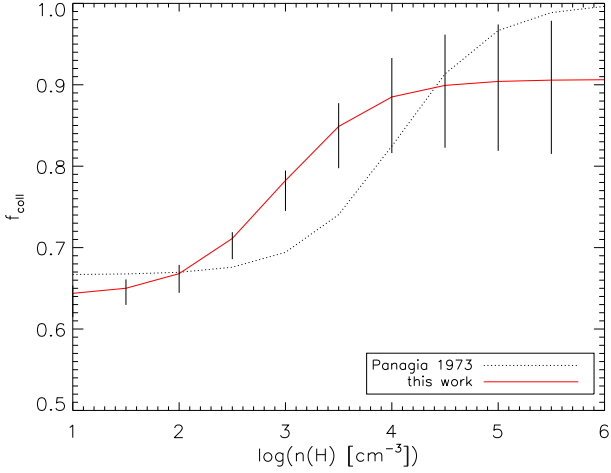,angle=90,width=\columnwidth}}
\caption{ $f_{\rm coll}$ factor accounting for the density effects. The red line represents the 3-parameter 
fit obtained on the basis of the photoionization modeling ($\tilde{f}_{\rm coll}$) to be
used in formula \ref{eq_lya_final} to calculate $L(\lya)$. At each density, vertical lines show the spread 
in  $\tilde{f}_{\rm coll}$ values
arising from the different T$_{bb}$ and $U$ models used: 
models with higher T$_{bb}$ and lower U have 
higher $\tilde{f}_{\rm coll}$ values.
The fit has been made for primordial nebular abundances and the 
black dotted line is $f_{\rm coll}$ from \citet{Panagia73}.}
\label{f_coll_comp}
\end{figure}

\subsubsection{Density effects}
Figure~\ref{all_prim} also shows that \lya\ is further enhanced at
high densities at the expense of two-photon emission, the sum of the
two luminosities being essentially constant if all other parameters
are kept fixed.
Collisional mixing of the relative populations of the $n=2$
levels between $2 ^2S$ and $2 ^2P$ of hydrogen alter the relative importance
of the rates of \lya\ and two-photon transitions resulting in increased line emission relative to the 2$\gamma$ continuum.

To correct for this effect under case~B assumptions,
\citet[see also \citeauthor{sti09} \citeyear{sti09}]{Panagia73} writes
$L(\lya) = \Qh \, \times h \nu_{\lya} \times f_{\rm coll}$ (cf.\ Eq.\ \ref{eq_lyacaseb}), where
\begin{equation}
  f_{\rm coll} \approx \frac{1 + a \times n_p}{1.5 + b \times n_p},
\label{eq_fcoll}
\end{equation}
with $a=b=1.35\times10^{-4}$, and where $n_p$ is the proton density. 
In this way the factor $f_{\rm coll}$ ranges from $2/3 \approx \alpha^{\rm eff}_{2 ^2P} / \alpha_B$ 
for low densities to $f_{\rm coll}=1$  for high densities. Under these assumptions, 
the two-photon continuum luminosity is $L(2\gamma)=\Qh \, \times h \nu_{\lya} \times (1-f_{\rm coll})$.

To derive the corresponding numerical factor from our detailed photoionization
models and to separate the density effect from the enhancement factor $P$ found above,
we define $\tilde{f}_{\rm coll}$ through:
\begin{equation}
  L(\lya) =  (1 - f_{\rm esc}) \Qh \, \times h \nu_{\lya} \times P \times \tilde{f}_{\rm coll}.
\label{eq_lya_final}
\end{equation}
The corresponding values of $\tilde{f}_{\rm coll}$, obtained for our grid of models for primordial 
nebulae are plotted in Figure~\ref{f_coll_comp} as a function of the input hydrogen density.
Also shown is, for comparison, Panagia's expression (Eq.\ \ref{eq_fcoll}) assuming $n_p=n_H$.
Notice that $\tilde{f}_{\rm coll}$ presented here is not derived from the \lya\ and 2$\gamma$ luminosities (which ranges from ~2/3 to 1 as in Panagia's formula)
but it is the fitting formula which has been calculated by comparing the right hand side of equation \ref{eq_lya_final} with our CLOUDY results.

For convenience, we derive the following 3-parameter fit:
\begin{equation}
\tilde{f}_{\rm coll} \approx \frac{1+ a \times n_H}{b + c \times n_H}.
\label{f_coll_fit}
\end{equation}
with $a=1.62 \times 10^{-3}$, $b=1.56$ and $c=1.78 \times 10^{-3}$. 
The coefficients obtained here differ from those of \citet{Panagia73} due to the fact that we have used a suite
of photoionization models covering a wide range of T$_{bb}$ and $U$ resulting in different T$_e$ 
and a different balance between processes in the nebula. However, the change in $\tilde{f}_{\rm coll}$ has little effect on the
values of $L(\lya)$ for low density. We use $\tilde{f}_{\rm coll}$ defined in this way for convenience since it allows us to reproduce the CLOUDY results.
As Fig.\ \ref{f_coll_comp} shows, our fit formula allows us to describe density
effects with an accuracy of $\pm 10 \%$ at high density and somewhat better
at low $n_H$.

The black dashed lines in Figure \ref{all_prim}, showing $P \times \tilde{f}_{\rm coll} / \tilde{f}_{\rm coll}(\rm log(n_H=1))$,
show how well our analytical prediction for \lya, including also density effects,
works at primordial metallicity.

\begin{figure}[htb]
\centerline{\psfig{figure=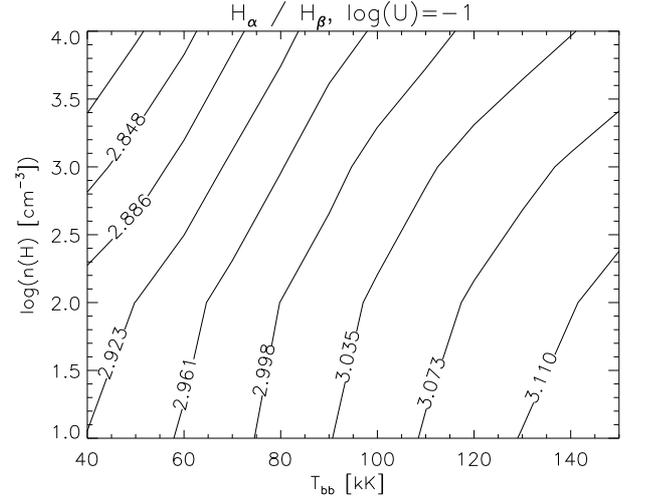,angle=90,width=\columnwidth}}
\caption{The Balmer decrement as a function of T$_{bb}$ and $n_H$ for primordial metallicity case. Note that H${\alpha}$ is
boosted for high T$_{bb}$ due to collisional excitation of hydrogen.}
\label{balmer_dec}
\end{figure}

\subsubsection{Applicable metallicity range}
The effect of collisional excitations, allowing for photoionization of H from 
excited states, depends on the electron temperature of the nebula,
and hence on its metallicity.
To find the range of nebular metallicity 
where our analytic expressions for $L(\lya)$ holds, we have computed model grids at different metallicities.
We can then ascertain how much our fit to the primordial case deviates from CLOUDY models as a function of metallicity.

The critical metallicity where the luminosity obtained by our fit deviates on average (for the entire grid of models) by 10 \% from
the correct one is $Z_{\rm coll} \approx 0.03$ \zsun. At higher metallicities,
cooler nebulae are produced and hence collisional effects play a smaller role in boosting hydrogen emission. However,
the effect only decreases gradually for higher metallicities
and depends on the detailed condition of the gas. 
For instance, in our simulations for Z$_{\rm{neb}}=0.05$ \zsun, the 
average deviation of the results obtained with our formula (primordial case) with respect to the CLOUDY results
is 13\%, while for  Z$_{\rm{neb}}=0.1$ \zsun\ it is 20\%.
The value of the ``transition'' metallicity $Z_{\rm coll} \approx 0.03$ \zsun\ below which
\lya\ (and the 2$\gamma$ continuum) are significantly boosted (modified) by departures
from case~B, should therefore only be taken as an indicative value. 
Tailored photoionization models are necessary for more accurate predictions.

\subsection{Other H lines}
The luminosities of other hydrogen lines are close to their case~B predictions as is shown in Figure~\ref{all_prim} for H${\alpha}$ (green symbols) and H${\beta}$ (blue symbols).
They are basically independent of $U$.

These lines are used to calculate the Balmer decrement which is conventionally used to measure the extinction.
The Balmer decrement (H${\alpha}$/H${\beta}$) obtained from our grid of models for $\log(U)=-1$ is shown
in Figure~\ref{balmer_dec}. It increases towards higher \tbb\ and exceeds the values commonly used, for example 2.86
for $n_e=10^2$ \cmc\ and $T_e=10$ kK (Dopita \& Sutherland 1996).
The reason for the enhanced H${\alpha}$/H${\beta}$ ratio is the collisional excitation of hydrogen, 
one of the contributors to enhanced \lya\ and 2$\gamma$ emission. This effect (see Osterbrock 2006)
occurs when photoelectrons carrying enough energy when colliding with H atoms, excite higher levels
followed by radiative cascade. Enhanced H$\alpha$ emission increases the Balmer decrement
which must be accounted for when calculating the extinction. An enhanced Balmer decrement was
already found, e.g.\ in tailored photoionization models for the metal-poor galaxy I Zw 18 
\citep{Davidson85,Stasinska99} and other giant \hii\ regions \citep[cf.][]{Luridiana03}, 
and has also been addressed by \citet{Luridiana09}.

\subsection{\heii\ lines}
\label{s_heii}
The \heii\ 1640~\AA\ line is of a particular interest since
it has been considered as one of the signatures of Pop~III or very metal-poor stars, which are expected
to have a very hard ionizing spectrum
emitting copious He$^+$ ionizing photons. Figure~\ref{all_prim} (orange symbols) shows \heii\ 1640 \AA\ emission 
line luminosities relative to their case~B values, calculated for the same $T_e$ and $n_e$ as the hydrogen lines. 
For lower ionization parameters, the line becomes weaker. This is due to an effect already discussed by
\citet{Stasinska86} in the context of planetary nebulae.
Photons with sufficient energy to ionize He$^+$ are also able to ionize hydrogen. When calculating the luminosity
of \heii\ in synthesis models, it is assumed that every photon with an energy $>$54 eV ionizes one He$^+$ ion.
The absorption of some 
high energy photons by H atoms results in a decrease in the number of photons available for producing \heii\
emission. The effect becomes significant at low ionization parameters since then the He$^+$/H$^0$ fraction decreases and
the probability of the high energy photon being absorbed by a H atom rather than by He$^+$ is higher.
This means that the synthesis models give an upper limit for the \heii\ luminosity.
We have not found a simple analytical prescription to account for the effect of this process on the \heii\ luminosity.

The behaviour of the equivalent width of \heii\ is discussed in Sect.\ \ref{s_sed}.

\begin{figure}[tbh]
\centerline{\psfig{figure=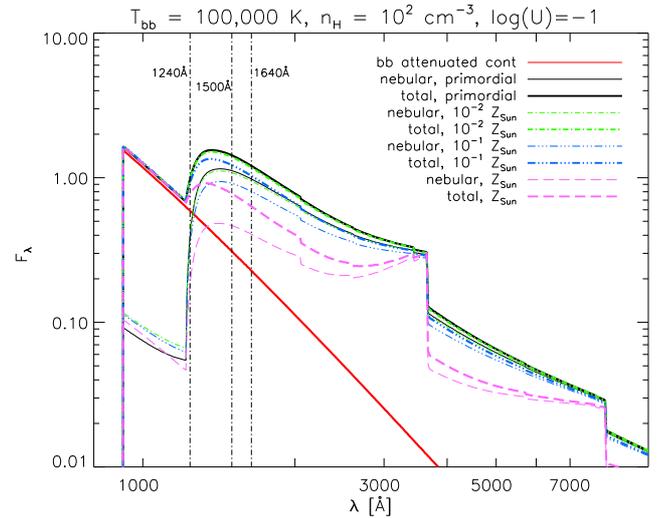,angle=90,height=7cm,width=\columnwidth}}
\caption{Continuum of T$_{bb}$=100,000 K models as a function of the nebular metallicity
for a constant density and ionization parameter.
The red solid line represents the transmitted stellar
(black body) continuum. The thin lines 
are nebular only contributions to the continuum.
The  thick lines are the total continuum emission (stellar + nebular). 
The first model is normalized at 1240 \AA\ and the rest are rescaled accordingly.}
\label{cont_zneb}
\end{figure}

\begin{figure}[tbh]
\centerline{\psfig{figure=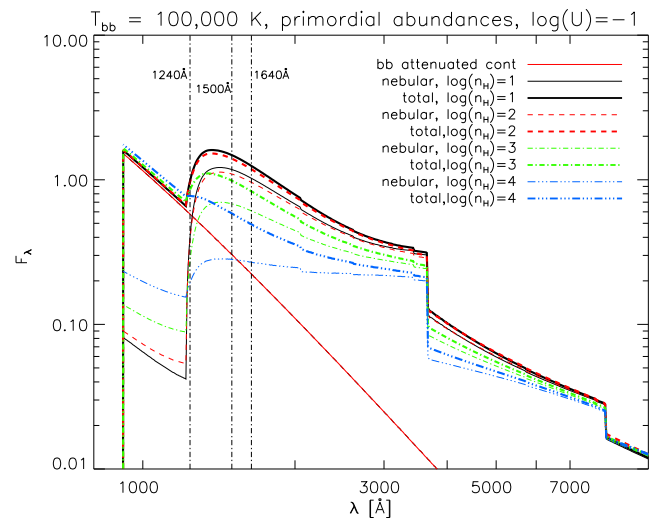,angle=90,height=7cm,width=\columnwidth}}
\caption{Continuum of T$_{bb}$=100,000 K models as a function of the hydrogen number density at the primordial
metallicity and constant ionization parameter. The red solid line represents the transmitted stellar
(black body) continuum. The thin lines 
are only nebular contributions to the continuum.
The thick lines are the total continuum emission (stellar + nebular). At higher densities 2$\gamma$
continuum gets destroyed due to collisions. First model normalized at 1240 \AA\ and the rest rescaled accordingly.}
\label{cont_hden}
\end{figure}

\subsection{Nebular continuum emission}
Nebular emission arising in \hii\ regions can contribute significantly
to the measured spectrum. The total nebular continuum is the sum of the free-free, free-bound and 2$\gamma$ 
continua of hydrogen and helium. Several nebular parameters affect its shape.
It depends on the electron temperature, ionization parameter, nebular metallicity and particle density
in the gas, as e.g.\ illustrated by Bottorff et al.\ (2006).
In Figure~\ref{cont_hden} we show the density dependence of the nebular continuum, which mainly affects the 
2$\gamma$ emission and hence also the shape of the continuum between \lya\ and the Balmer jump.
Figure~\ref{cont_zneb} shows the dependence on the nebular metallicity. For higher metallicities, the increased 
cooling lowers the total emission in the nebular continuum, and its shape is altered 
as expected from atomic physics (see Bottorff et al.\ 2006) due to the decrease of the average
electron temperature at higher metallicity.
Evolutionary synthesis models such as ours, cannot properly describe the variety of 
shapes and strengths of the nebular continuum shown here, since they rely on 
simple assumptions such as constant nebular density and temperature. 
Again, tailored photoionization models are necessary for more accurate predictions.

\section{Photoionization models for realistic SEDs}
\label{s_sed}
In this Section we show how to relate realistic SEDs obtained e.g.\ from
evolutionary synthesis models with the results from the photoionization
models discussed above, which used black body spectra.
Finally, we show the updated predictions for the equivalent widths
of \lya\ and \Heiiuv\ lines obtained from our CLOUDY models.

\begin{figure*}[tbh]
\centerline{\psfig{figure=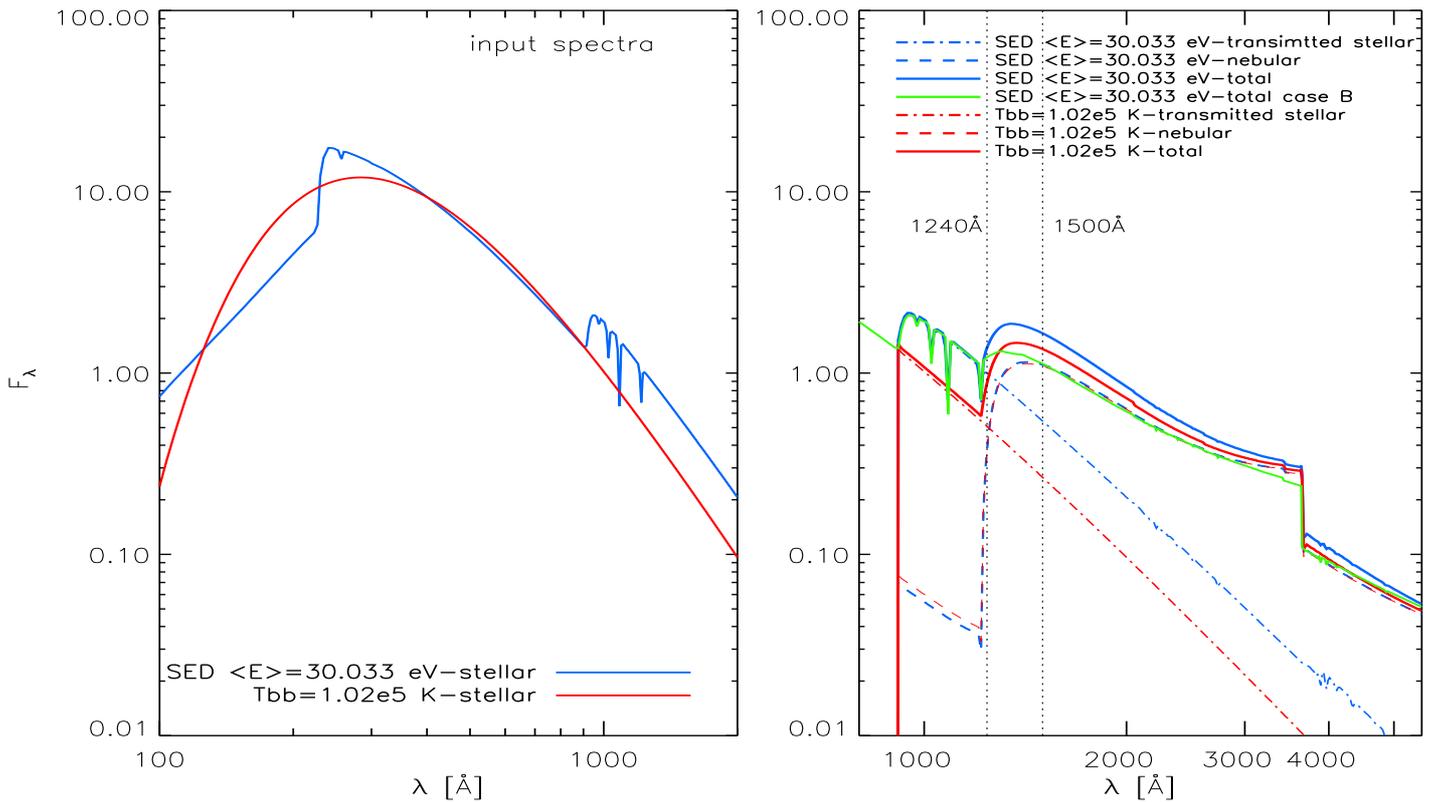,angle=90,height=12cm,width=20cm}}
\caption{{\bf Left:} Comparison of input stellar spectra for photoionization models showing a realistic SED 
(blue line; Pop~III, Salpeter IMF 1-100 M$_{\odot}$ at zero age)
and the corresponding black body spectrum (red line) with the same average photon energy 
($\overline E=30.033$ eV) in the Lyman continuum.
The SED model is normalized at 1240 \AA\  and the rest
rescaled to match the same ionizing flux $Q(H)$.
{\bf Right:} Input stellar spectra (dash-dotted, same colours as left panel) 
and predicted nebular and total continua
from our CLOUDY models for $\log(n(\rm H)) = 1$ \cmc\ and  $\log(U) = -1$ (solid lines, using the same colours as in left panel;
cf.\ also inset for symbols). 
The green solid line shows the total (stellar + nebular) SED from our
evolutionary synthesis model assuming constant nebular density and temperature and 
case~B. Notice the difference between case~B and the CLOUDY result due to enhanced 2$\gamma$ emission.}
\label{cont_comp}
\end{figure*}

\subsection{How to connect realistic SEDs with black body calculations}

\begin{figure*}[tbh!]
\centering
\includegraphics[height=20cm,angle=90]{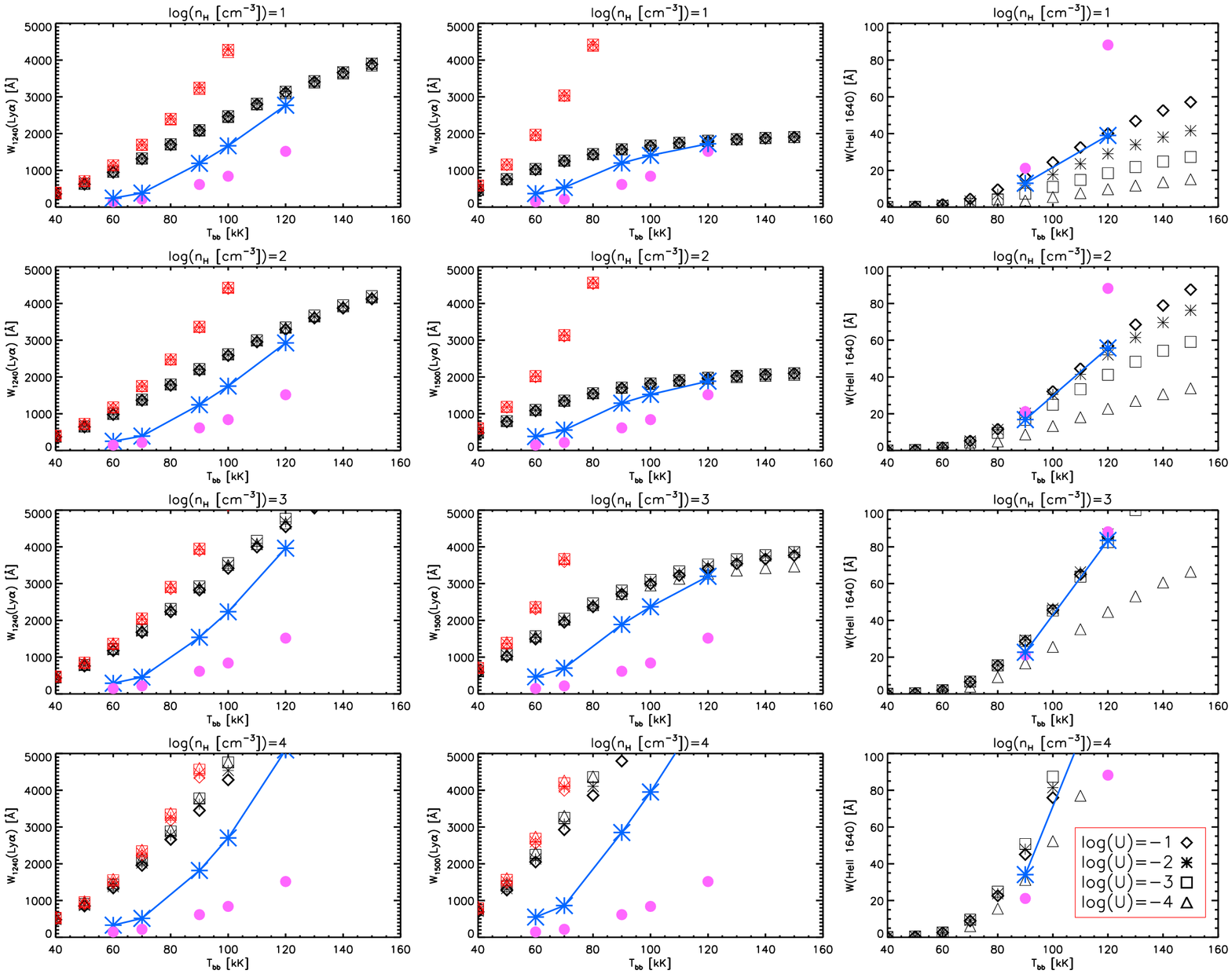}
\caption{Equivalent widths ($W_{\lambda}$) of \lya\ at 1240 (column 1) and 1500 \protect\AA\ (column 2) and \Heiiuv\ (column 3)
for a primordial metallicity nebula at different densities and ionization
parameters (marked with different symbols as defined in the inset). The red symbols show an upper limit for $W_{\lambda}$(\lya)
since they are calculated using only stellar (black body) continuum. 
Blue symbols -- the most realistic predictions -- show
the results obtained for the corresponding stellar population SEDs (models with the same $P$ in the case of \lya\ and
with the same $\log(\Qhep/\Qh)$ for the \heii\ analysis, for simplicity only models with $\log(U) = -1$ are shown) and the pink ones their
case~B predictions. Notice the dependence of the \Heiiuv\ equivalent width on the
ionization parameter and the discrepancy between its case~B and photoionization model predictions for real SEDs.
}
\label{ew}
\end{figure*}

Since the number of H ionizing photons and their mean energy are the main parameters 
determining the luminosity of the H recombination lines and of the 2$\gamma$ continuum,
$Q(H)$ and $P$ suffice to compute these quantities from arbitrary SEDs. 
To account also for density effects, 
Eqs.\ \ref{eq_lya_final} and  \ref{f_coll_fit} can be used to yield the correct
\lya\ luminosity. Numerical tests using SEDs described in Sect.\ \ref{s_uv_synthesis}
show the validity of this result, typically within 2-3 \%.
For black bodies the relation between $P$ and \tbb\ is given by
\begin{equation}
 \tbb[\rm kK] = -53.42 + 69.85 \, P
\label{eq_p_bb}
\end{equation}
to a good approximation.
The black body temperature $\tbb(P)$ corresponding to the more realistic
SED is thus easily determined.

For other quantities, such as the \heii\ line luminosity, the correspondence
between black bodies and other SEDs is different. For example, since the 
relative line ratios of \heii/H lines depend to first order on the relative
number of He$^+$/H ionizing photons, the black body with the same 
hardness \Qhep/\Qh\ is the most appropriate.
For black body spectra one has, to a good approximation:
\begin{eqnarray}
\lefteqn{ \tbb[\rm kK]=314.6 + 382.4\,x + 268.7\,x^2 +{}}
\nonumber \\
& & {} + 103.3\,x^3 + 20.12\,x^4 + 1.545\,x^5. {}
\label{eq_tx}
\end{eqnarray}
where $x = \log(\Qhep/\Qh)$.
Again, we have tested a number of models and confirmed that one can get the same 
luminosity (within a few \%) using SEDs and corresponding black body models,
for otherwise identical nebular parameters.
Differences of $\sim 10$\% can appear in case of significantly (several eV) different values 
of $P$. In principle not only the stellar \Qhep/\Qh\ ratio determines the \heii\ emission, 
but also the conditions in the gas. Since $P$ establishes the electron temperature, 
significant differences in $P$ result in different luminosity coefficients (recombination
rates and emissivities).

In Fig.\ \ref{cont_comp} we show such a comparison between a realistic SED from our synthesis models
(Pop~III, Salpeter IMF 1--100 M$_{\odot}$ at zero age, instantaneous burst) and the appropriate black body, chosen such as to 
reproduce the correct H line luminosities and nebular continuum. 
Here the average photon energy in the Lyman continuum is $\overline E=30.033$ eV, hence $\tbb=102$ kK.
The input spectra, shown on the left panel, are scaled to the same total Lyman continuum
flux $Q(H)$. The nebular parameters are $\log(n_H) = 1$ \cmc, $\log(U)=-1$, and primordial composition.

Several interesting points are illustrated with this figure.
First, as the left panel shows, it must be remembered that realistic SEDs show a 
Lyman break due to the hydrogen opacity in the atmosphere of the hot stars 
responsible for the flux. For young ages this break corresponds typically to a 
flux increase by $\sim$ 0.2-0.3 dex at 912 \AA, as discussed e.g.\ by 
\citet{Scha03}\footnote{The amplitudes of the Lyman break, 912$^+$/912$^-$, given in 
\citet{Scha03} for very low metallicities are correct, but misleading.
Indeed, with the definition adopted there, the 912$^+$ flux average 
(over 1080--1200 \AA) includes strong absorption from the \ion{He}{ii} $\lambda$1084 line.
This leads to values of 912$^+$/912$^-$ $< 1$ despite the fact that
all models show a Lyman break in absorption.}.
In consequence, calculations relying on black body spectra underestimate
the observable UV continuum, as shown in the right panel of Fig.\ \ref{cont_comp}, 
leading to non-negligible differences e.g.\ for equivalent widths predictions 
of emission lines (cf.\ below).
In Fig.\ \ref{cont_comp} (right panel) we also show the stellar + nebular SED
predicted by our synthesis models assuming constant nebular density and temperature and 
case~B (green line). The differences between this and the full CLOUDY model are mainly due to the boosting of the 
2$\gamma$ continuum discussed earlier, and to varying $n_e$ and $T_e$.

\subsection{\lya\ equivalent width predictions}
The measurement of the equivalent width of \lya\ is both observationally and interpretationally difficult, 
since the continuum around it is affected on one side at high redshifts by \lya\
forest absorption and on the other by an unknown combination of 2$\gamma$ continuum and starlight.
Furthermore, when not measured from spectroscopy, it is common practice to measure $W(\lya)$ 
via a line to continuum ratio with the continuum estimated at another 
(usually longer) wavelength $\lambda_{\rm cont}$, e.g. $W_{\lambda}=L_{\rm line}/ F_{\lambda_{\rm cont}}$.
Different methods to measure $W(\lya)$ have been reviewed by \citet{Hayes06}.
To illustrate the possible impact or uncertainty related to the way the continuum is estimated 
we subsequently plot \lya\ equivalent widths predicted from the CLOUDY models
using the continuum at both 1240 and 1500~\AA\ as a reference.

Figure~\ref{ew} shows the predictions for $W(\lya)$ from all CLOUDY models 
for primordial composition, as a function of \tbb, ionization parameter and hydrogen density.
Black symbols show the predictions using black body spectra and accounting for
the nebular continuum; red symbols the same but neglecting the nebular continuum.
Blue symbols show the results from CLOUDY models using SEDs from our
Pop~III synthesis models for zero age populations, plotted at the corresponding
$\tbb(P)$ value (see above). Pink symbols show $W(\lya)$ predicted from 
our standard synthesis models neglecting the boost of \lya\ (case~B departure).

\begin{figure}[htb!]
\centering
\includegraphics[height=10cm,angle=90,width=\columnwidth]{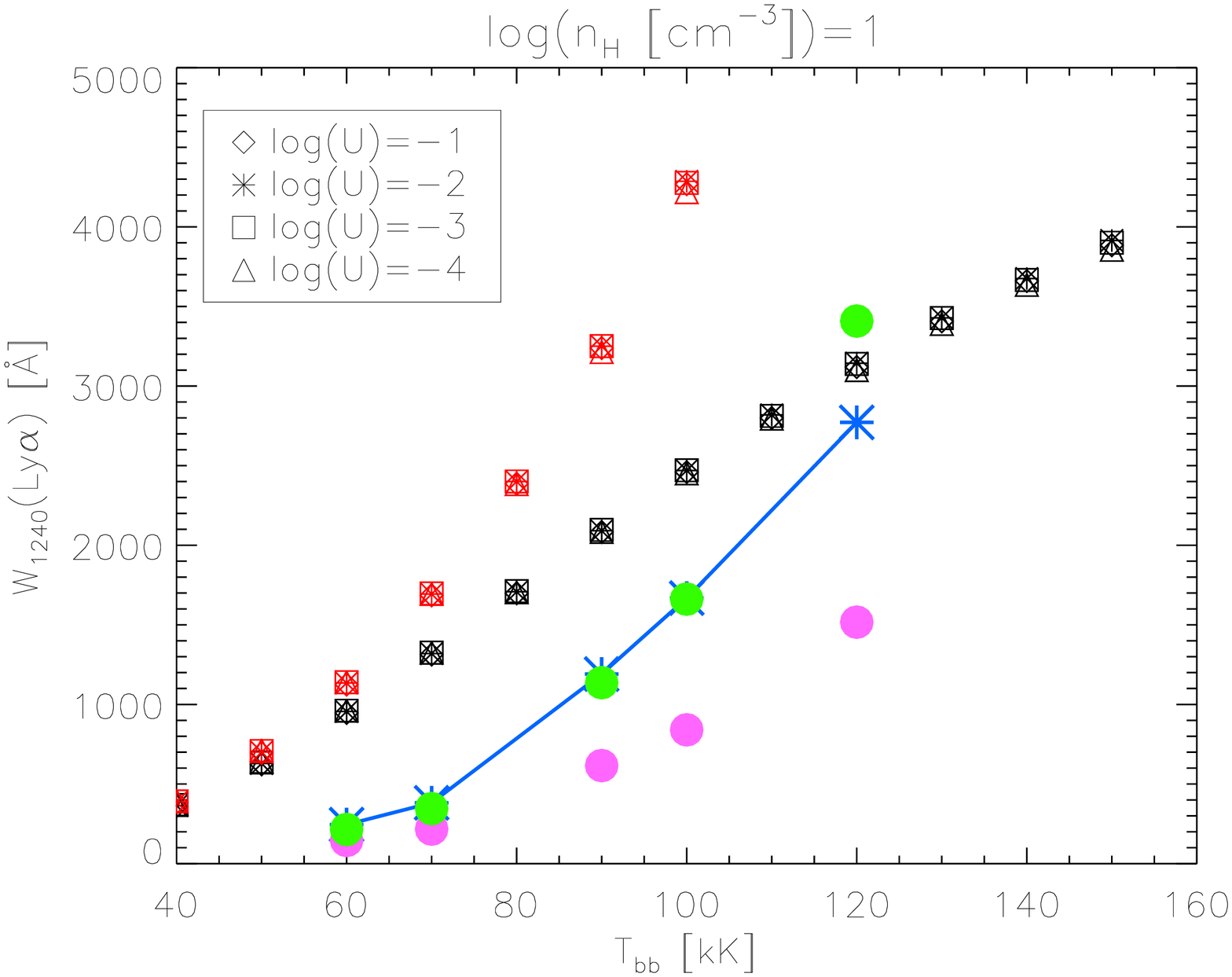}
\caption{Equivalent widths ($W_{\lambda}$) of \lya\ at 1240 \protect\AA\ (first column, second row of Fig.\ \ref{ew}).
All the symbols as in Fig.\ \ref{ew}. Additionally, the green symbols show the case~B SED 
results (pink symbols) corrected using our formula (first order correction; only the enhancement
in the line flux is taken into account, there is no correction for the continuum).
The difference coming from not correcting for the nebular continuum enhancement is
visible for the hottest model.}
\label{ew_lya_only}
\end{figure}

Overall $W(\lya)$ increases with the \tbb\ or equivalent since the \lya\
luminosity ($\propto Q(H) \times P$) increases more rapidly than the continuum
flux close to \lya. For very hot models and if 1500 \AA\ is taken as a reference 
for the continuum, $W(\lya)$ tends to a maximum value (here $\sim$ 2000 \AA,
as shown in col.\ 2) since line emission and the dominating nebular continuum 
scale in the same manner.
For the reasons discussed above (absence of the Lyman break), predictions based on 
black body spectra (black and red symbols) overestimate $W(\lya)$ compared to models 
including more realistic stellar SEDs (in blue).
Differences between black body models including or neglecting the nebular 
continuum (black and red) decrease with increasing ISM density,
because of the decrease of the 2$\gamma$ continuum. This also explains
the decreasing difference between $W(\lya)$ using 1240 and 1500~\AA\ as a 
reference, when $n_H$ increases.
Last, but not least, $W(\lya)$ predicted by our standard synthesis models 
for the ZAMS (pink symbols) fall below the more accurate CLOUDY predictions
(blue), since the latter account for case~B departures. Once applying the simple
correction given by Eq.\ \ref{eq_caseb_corr} one obtains the
result which takes into account the increase in \lya\ line flux. However, that is only
the first order correction which does not include the difference coming from 
the continuum, mainly a similar increase in the 2$\gamma$ continuum.
In cases when the continuum is measured around \lya\ (1240 \AA) stellar flux is dominant 
so the correction described above gives a good estimate of the correct $W(\lya)$ value (see Fig. \ref{ew_lya_only}).

\subsection{\Heiiuv\ equivalent width}

The right column in Fig.\ \ref{ew} shows the predicted \Heiiuv\ equivalent widths for the same
CLOUDY models except that $\tbb(x)$ from Eq.\ \ref{eq_tx} is 
used here . The nebular continuum predicted at 1640 \AA\ is included in all calculations of $W(1640)$.
In contrast to \lya, the \heii\ equivalent width depends on the ionization parameter, since
the line luminosity changes due to the ``Stasinska-Tylenda effect''. 
For high ISM densities $W(1640)$ increases due to the decrease of the nebular continuum.
Differences between models assuming black body spectra (black symbols) and realistic SEDs
(blue) are minor for $W(1640)$ as long the nebular continuum dominates at these
wavelengths. This is true for sufficiently hot spectra ($\gtrsim$ 80 kK).
Compared to predictions from our standard evolutionary synthesis models (pink symbols)
the \Heiiuv\ equivalent widths predicted by CLOUDY are significantly lower for 
$\tbb \ga 100$ kK. This is due to two effects, the Stasinska-Tylenda effect reducing the line flux,
and the increase of the 2$\gamma$ nebular continuum.
At lower \tbb, case~B predictions give an upper limit (corresponding to our highest ionization parameter case)
for  \Heiiuv.
Only for a high ionization parameter and/or high density do photoionization models
predict $W(1640)$ values as high as those given by \citet{Scha03}.

\subsection{Summary: correction of evolutionary synthesis model results
for departures from case B}
In this section we give a recommended step-by-step procedure which allows the correction of the
synthesis models for the departures from case~B in ionization bounded regions
(i.e.\ with no leakage of Lyman continuum photons), relevant especially for the case of very hot stars
and very low metallicities.

\noindent For L(Lya):
\begin{itemize}
\item Take the SED (function of age, IMF, ...).
\item Get (or compute) the mean energy of Lyman continuum photons, $P$, in units of 13.6 eV.
\item Compute the ``collisional'' factor $\tilde{f}_{\rm coll}$ from our fit formula (Eq.\ \ref{f_coll_fit}).
\item Multiply the synthesis model result $L(\lya)$ by $P*\tilde{f}_{\rm coll}/(2/3)$
  (Eq.\ \ref{eq_caseb_corr}).
\end{itemize}

\noindent For W(\lya):
divide the equivalent width obtained by the synthesis model of interest by 2/3 and multiply it by $P*\tilde{f}_{\rm coll}$.
This provides the first order correction for $W(\lya)$, neglecting the increase of the nearby nebular continuum.
An example following this procedure is shown by the green symbols  in Fig.\ \ref{ew_lya_only}.

For \heii\ recombination lines:
no simple correction is possible. Synthesis models give an upper limit for the luminosity, an upper
limit for equivalent width up to log$(\Qhep/\Qh) \sim -1.2$ for the primordial metallicity case and overestimate it for the highest 
\Qhep/\Qh\ cases (low density) due to the lack of accounting for enhanced 2$\gamma$ emission.

\section{Discussion and implications}
\label{s_discuss}

\subsection{Dependence on model assumptions}
\label{s_dependence}
All our CLOUDY models have been computed for ionization bounded nebulae corresponding
to no escape  of Lyman continuum photons, i.e.\ $f_{\rm esc}=0$.  
For density (or matter) bounded cases with $f_{\rm esc}>0$, quantities such as
hydrogen recombination line luminosities and the luminosity of the nebular continuum
longward of \lya\ can be rescaled  to first order by scaling with $(1- f_{\rm esc})$, as already introduced 
in Eq. \ref{eq_lya_final}. Such a scaling is appropriate since the corresponding emissivities are
approximately constant across the nebula. However, since He$^+$ recombination
lines (if present) are emitted only in the innermost parts of the nebula,  \heii\ line luminosities
will not be reduced in density bounded regions, as long the radius of the He$^+$ sphere
remains smaller than that of density bounded  \hii\ region.
These behaviours are also expected in non-spherical regions, as demonstrated e.g.\ by
\citet{Johnson09}.
In consequence, both $L(\lya)$ and $W(\lya)$ will decrease with increasing Lyman continuum
escape (although the latter not proportionally with $f_{\rm esc}$), whereas the opposite will be true 
for $W($\Heiiuv$)$ and for other \heii\ lines, where the underlying continuum is mostly of nebular origin.
In other words, the decrease of $W($\Heiiuv$)$ found in this paper due to competition with H ionizing
photons, may be mitigated in  objects with significant leakage of Lyman continuum radiation, leading
again to higher  \heii\ equivalent widths \citep[cf.][]{Scha02,Scha03}.
For other effects of ``leaking'', i.e.\ star-forming galaxies with density bounded \hii\ regions, see the 
photoionization models of \citet{Inoue10}.

We shall now discuss some implications of our modeling for the interpretation of the hydrogen and helium emission lines and the nebular continuum observed in high redshift emission line galaxies.

\subsection{Lyman continuum output}
As metallicity decreases, the Lyman continuum output of the stellar
population, both relative to the observed stellar UV continuum longward of
\lya\ and per unit stellar mass, increases. The amount of this increase 
depends obviously also on the detailed shape of the IMF. 
For example, for a Salpeter IMF with a uniform upper mass cut-off of
100 \msun, the ionizing photon rate per unit UV luminosity,
$Q_H/L_{1500}$, increases by a factor 3 approximately between solar
an zero metallicity. For the most extreme IMF considered here 
this increase is up to a factor 10 (see Fig.\ \ref{f_nlyc}).
The effect of this
is to increase the line and nebular continuum emission relative to any
observable stellar continuum. In extreme cases \citep[e.g.,][]{Raiter2010}
the nebular emission may completely dominate the UV/visible/NIR
spectrum. 
Also, if low metallicity objects and/or different IMFs are relevant for the sources of cosmic
 reionization, their intrinsic Lyman continuum flux, generally estimated from UV restframe observations, 
may need to be revised accordingly.

\subsection{SFR(UV)}
The estimation of SFR from the level of the UV continuum \citep[cf.][]{Kenn98} is based on evolutionary synthesis
models assuming ``standard'' stellar populations and IMFs.
At metallicities below solar, the output of observable UV continuum radiation (i.e.\ typically
at 1500 \AA\ restframe) per unit stellar mass increases somewhat, due to the increasing average 
temperature of the stars (cf.\ Fig.\ \ref{f_sfr}).
A maximum in $L_{1500}$ is typically reached at $Z/z\sun \sim 1/100$. Below this,
the average spectrum shifts more strongly into the Lyman continuum, leading to a
decrease of stellar radiation in the observable UV, which is, however, 
compensated by increasing nebular emission. The net result, illustrated in Fig.\ \ref{f_sfr}
for constant SFR, is that star-formation rates derived from the UV may need to be revised downward
at low metallicities with respect to the ``standard'' calibration of \citet{Kenn98},
but typically by less than a factor 2.

\subsection{\lya}
One of the main findings of our study is the enhancement of \lya\ emission
at low metallicities with respect to the commonly used case B value.
This departure from case B occurs even at moderaly low metallicity, say
$Z/\zsun \la 1/10$, and becomes more important for lower $Z$.
In consequence, higher intrinsic \lya\ line luminosities relative to the UV continuum (to $L_{\rm bol}$, SFR, or to another
measure of the rate of massive star-formation) are expected and higher \lya\ equivalent widths.
Even higher values of $W(\lya)$ can be obtained in regions with a high ISM
density, where \lya\ emission can further be ``boosted'' at the expense of
the 2-photon nebular continuum (see Fig.\ \ref{all_prim}).

With respect to our earlier predictions \citep[cf.][]{Scha03}, $W(\lya)$ is increased
by a factor 2--2.5 for Pop III objects, and by more than 70\% at moderately
low metallicity  ($Z/\zsun \la 1/20$, cf.\ Fig.\ \ref{f_wlya}), for a given IMF.
This increase of \lya\ may help to understand  objects with large \lya\ equivalent
widths found in some surveys,  without the need for recourse to unusual IMFs or to a clumpy 
ISM \citep[cf.][]{Malhotra02,Daws04,Fin08}. 

Finally, the enhanced \lya\ emission found here should represent good news for
searches for very distant/early galaxies, since the intrinsic \lya\ emission
of metal-poor objects is shown to be considerably higher than previously thought,
reaching up to 20--40\% of $L_{\rm bol}$ in Pop~III dominated objects.
Of course, the intrinsic \lya\ emission (and hence also $W(\lya)$) will be lower
in objects, having a non-zero escape of flux from the Lyman continuum.
Furthermore, several processes exist (e.g.\ dust absorption, scattering out of the 
line-of-sight in the ISM and/or IGM), which will reduce the \lya\ 
emission on the way from the source to the observer.

\subsection{\heii\  emission}
The reduced \heii\ emission found in this paper, compared to our earlier predictions assuming simple
case B recombination, may indicate that ongoing searches for signatures of Pop~III stars using this feature
\citep[see e.g.\ review by][]{Schaerer08} could be more difficult than anticipated so far.
The corresponding upper limits of Pop~III star-formation rate density derived e.g.\ from the survey
of \citet{Nagao08} need then to be revised upwards.

The detection of the \Heiiuv\ line indicates a significant number of high energy photons in the ionizing spectrum. This line has therefore been sought,
but not detected so far at high-$z$ (Nagao 2008). Prescott et al. (2009) have detected this line from a $z$=1.67 spatially extended nebula (or \lya\ blob)
but have not been able to determine the nature of the ionizing source.
We find that the line (and therefore the equivalent width) could be fainter than expected previously, due to its dependence on the ionization parameter.
This suggests that deeper observations will be needed to detect it.
Additionally, another difficulty comes from the fact that for the hottest (the highest $\Qhep/\Qh)$ ratio) stars the strongest enhancement in 2$\gamma$
continuum is predicted which further decreases the measured $W(1640)$.
However, if Pop~III stars are present in objects with significant escape fractions 
in the Lyman continuum, the \Heiiuv\ equivalent width may again be stronger, due to
the reduced nebular continuum (cf.\ Sect.\ \ref{s_dependence}).

\subsection{Nebular continuous emission}
The overall nebular continuum emission is dependent on several nebular parameters
and can result in a variety of spectral shapes (Fig.\ \ref{cont_zneb} and  \ \ref{cont_hden}).
Note that our simulations have been carried out in the absence of dust which would add another 
parameter affecting the shape of the continuum.

Proper photoionization modeling shows  the importance of the 2$\gamma$ continuum produced in a nebula ionized by very hot stars.
At low ISM density and primordial/low metallicity, it is boosted in the same way as \lya\ and can completely dominate the nebular emission at 1216--1600 \AA.
This can affect the broadband flux measurements and reduce the equivalent widths of UV emission lines, particularly \heii.
The strength of the nebular continuum and the expected variations of its spectral shape
also indicate that measurements of the \lya\ equivalent width from photometry may
be more uncertain than naively expected \citep[cf.][]{Hayes06}.

As can be seen from Fig.\ \ref{cont_comp}, the slope of the UV continuum 
between $\sim$ 1300 -- 2000 \AA\ obtained from our CLOUDY model is very similar to that
predicted by the evolutionary synthesis models assuming the same case~B assumption, constant
electron density and temperature.  When the 2$\gamma$ continuum is included, one 
cannot obtain significantly steeper slopes than found for this Pop~III simulation, since the electron 
temperature which affects the shape of 2$\gamma$  emissivity \citep[cf.][]{Bottorff06} becomes shallower
as metallicity increases.

\section{Conclusions}
\label{s_conclude}
Building on the earlier calculations of \citet{Scha02,Scha03}, we have computed evolutionary synthesis
models for a wide range of metallicities from Pop~III (zero metallicity) to solar metallicity covering
a wider range of IMFs -- including power laws and log-normal IMFs with different characteristic
masses -- than before.
Using these synthesis models, we present the expected UV properties of star-forming galaxies,
including their  Lyman continuum fluxes, UV  luminosity, and properties of the continuum
(e.g.\ $\beta$-slopes), as well as properties of important emission lines such as \lya\ 
and \Heiiuv\ (see Sect.\ \ref{s_uv_synthesis}).

To investigate possible departures from the simple case~B recombination theory assumed in many synthesis
models, including ours, we have computed large grids of CLOUDY photoionization models
for zero and low metallicity nebulae, using both black-body spectra and the SEDs predicted by our
synthesis models (Sects.\ \ref{s_cloudy} and \ref{s_sed}).

Our main conclusions from the photoionization models are the following:
\begin{itemize}
\item Due to departures from case~B (collisional excitation and ionization from excited levels),
both \lya\  and 2$\gamma$ continuum emission can be significantly enhanced at low nebular densities.
Their strengths are found to scale with the mean photon energy of the ionization source  in the Lyman continuum.
\item The equivalent width of \lya\ can be larger than expected from case~B calculations due to the line flux 
enhancement. The measurement can also be affected by 2$\gamma$ emission if the continuum is measured at longer
wavelength.
\item Further enhancement of \lya\ at the expense of 2$\gamma$ emission can occur due to collisional 
mixing between the hydrogen $2 ^2S$--$2 ^2P$ levels at higher densities.
\item \heii\ emission line fluxes (and consequently their equivalent widths)
can be significantly decreased due to their dependence on the ionization parameter.
This could make searches for the \Heiiuv\ line at high-$z$ more difficult.
\item The enhancement of the 2$\gamma$ continuum and its dependence on nebular parameters
can result in reduced equivalent widths of the UV emission lines and also change the UV restframe 
colours of high-$z$ galaxies.

\end{itemize}

Our results are of relevance to searches for, and interpretation of observations of, 
metal-poor and/or high-$z$ galaxies which may host the first stars that appeared 
in the Universe. 

\begin{acknowledgements}
We thank Suzy Collin, Mike Fall, and Grazyna Stasi\'nska
for helpful and stimulating discussions.
Some explorations of photoionization models were already undertaken
earlier with Francois Ricquebourg and Michael Zamo. DS wishes to thank
them here for their contributions. 
The work of DS is supported by the Swiss National Science Foundation.
DS is thankful for support from the ESO visitor program,
during which part of this work was done. 
\end{acknowledgements}

\bibliographystyle{aa} 
\bibliography{references}   

\begin{thebibliography}{49}
\expandafter\ifx\csname natexlab\endcsname\relax\def\natexlab#1{#1}\fi

\bibitem[{{Balestra} {et~al.}(2010){Balestra}, {Mainieri}, {Popesso},
  {Dickinson}, {Nonino}, {Rosati}, {Teimoorinia}, {Vanzella}, {Cristiani},
  {Cesarsky}, {Fosbury}, {Kuntschner}, \& {Rettura}}]{Balestra10}
{Balestra}, I., {Mainieri}, V., {Popesso}, P., {et~al.} 2010, \aap, 512, A12+

\bibitem[{{Bottorff} {et~al.}(2006){Bottorff}, {Ferland}, \&
  {Straley}}]{Bottorff06}
{Bottorff}, M.~C., {Ferland}, G.~J., \& {Straley}, J.~P. 2006, \pasp, 118, 1176

\bibitem[{{Bouwens} {et~al.}(2010){Bouwens}, {Illingworth}, {Oesch}, {Trenti},
  {Stiavelli}, {Carollo}, {Franx}, {van Dokkum}, {Labb{\'e}}, \&
  {Magee}}]{Bouwens10_betaz7}
{Bouwens}, R.~J., {Illingworth}, G.~D., {Oesch}, P.~A., {et~al.} 2010, \apjl,
  708, L69

\bibitem[{{Bromm} {et~al.}(2001){Bromm}, {Kudritzki}, \& {Loeb}}]{Bromm01}
{Bromm}, V., {Kudritzki}, R.~P., \& {Loeb}, A. 2001, \apj, 552, 464

\bibitem[{{Ciardi} {et~al.}(2001){Ciardi}, {Ferrara}, {Marri}, \&
  {Raimondo}}]{Ciardi2001}
{Ciardi}, B., {Ferrara}, A., {Marri}, S., \& {Raimondo}, G. 2001, \mnras, 324,
  381

\bibitem[{{Davidson} \& {Kinman}(1985)}]{Davidson85}
{Davidson}, K. \& {Kinman}, T.~D. 1985, \apjs, 58, 321

\bibitem[{{Dawson} {et~al.}(2004){Dawson}, {Rhoads}, {Malhotra}, {Stern},
  {Dey}, {Spinrad}, {Jannuzi}, {Wang}, \& {Landes}}]{Daws04}
{Dawson}, S., {Rhoads}, J.~E., {Malhotra}, S., {et~al.} 2004, \apj, 617, 707

\bibitem[{{Ferland} {et~al.}(1998){Ferland}, {Korista}, {Verner}, {Ferguson},
  {Kingdon}, \& {Verner}}]{Ferland98}
{Ferland}, G.~J., {Korista}, K.~T., {Verner}, D.~A., {et~al.} 1998, \pasp, 110,
  761

\bibitem[{{Finkelstein} {et~al.}(2010){Finkelstein}, {Papovich}, {Giavalisco},
  {Reddy}, {Ferguson}, {Koekemoer}, \& {Dickinson}}]{Fin10}
{Finkelstein}, S.~L., {Papovich}, C., {Giavalisco}, M., {et~al.} 2010, \apj,
  719, 1250

\bibitem[{{Finkelstein} {et~al.}(2008){Finkelstein}, {Rhoads}, {Malhotra},
  {Grogin}, \& {Wang}}]{Fin08}
{Finkelstein}, S.~L., {Rhoads}, J.~E., {Malhotra}, S., {Grogin}, N., \& {Wang},
  J. 2008, \apj, 678, 655

\bibitem[{{Gnedin} {et~al.}(2008){Gnedin}, {Kravtsov}, \& {Chen}}]{Gnedin08}
{Gnedin}, N.~Y., {Kravtsov}, A.~V., \& {Chen}, H. 2008, \apj, 672, 765

\bibitem[{{Hayes} \& {{\"O}stlin}(2006)}]{Hayes06}
{Hayes}, M. \& {{\"O}stlin}, G. 2006, \aap, 460, 681

\bibitem[{{Inoue}(2010)}]{Inoue10}
{Inoue}, A.~K. 2010, \mnras, 401, 1325

\bibitem[{{Johnson} {et~al.}(2009){Johnson}, {Greif}, {Bromm}, {Klessen}, \&
  {Ippolito}}]{Johnson09}
{Johnson}, J.~L., {Greif}, T.~H., {Bromm}, V., {Klessen}, R.~S., \& {Ippolito},
  J. 2009, \mnras, 399, 37

\bibitem[{{Kennicutt}(1998)}]{Kenn98}
{Kennicutt}, Jr., R.~C. 1998, \araa, 36, 189

\bibitem[{{Larson}(1998)}]{Larson98}
{Larson}, R.~B. 1998, \mnras, 301, 569

\bibitem[{{Luridiana}(2009)}]{Luridiana09}
{Luridiana}, V. 2009, \apss, 324, 361

\bibitem[{{Luridiana} {et~al.}(2003){Luridiana}, {Peimbert}, {Peimbert}, \&
  {Cervi{\~n}o}}]{Luridiana03}
{Luridiana}, V., {Peimbert}, A., {Peimbert}, M., \& {Cervi{\~n}o}, M. 2003,
  \apj, 592, 846

\bibitem[{{Malhotra} \& {Rhoads}(2002)}]{Malhotra02}
{Malhotra}, S. \& {Rhoads}, J.~E. 2002, \apjl, 565, L71

\bibitem[{{Nagao} {et~al.}(2008){Nagao}, {Sasaki}, {Maiolino}, {Grady},
  {Kashikawa}, {Ly}, {Malkan}, {Motohara}, {Murayama}, {Schaerer}, {Shioya}, \&
  {Taniguchi}}]{Nagao08}
{Nagao}, T., {Sasaki}, S.~S., {Maiolino}, R., {et~al.} 2008, \apj, 680, 100

\bibitem[{{Osterbrock} \& {Ferland}(2006)}]{Osterbrock06}
{Osterbrock}, D.~E. \& {Ferland}, G.~J. 2006, {Astrophysics of gaseous nebulae
  and active galactic nuclei}

\bibitem[{{Panagia}(1973)}]{Panagia73}
{Panagia}, N. 1973, \aj, 78, 929

\bibitem[{{Panagia}(2002)}]{Panagia02}
{Panagia}, N. 2002, ArXiv Astrophysics e-prints

\bibitem[{{Panagia}(2005)}]{Panagia05}
{Panagia}, N. 2005, in Astrophysics and Space Science Library, Vol. 327, The
  Initial Mass Function 50 Years Later, ed. {E.~Corbelli, F.~Palla, \&
  H.~Zinnecker}, 479--+

\bibitem[{{Popesso} {et~al.}(2009){Popesso}, {Dickinson}, {Nonino}, {Vanzella},
  {Daddi}, {Fosbury}, {Kuntschner}, {Mainieri}, {Cristiani}, {Cesarsky},
  {Giavalisco}, {Renzini}, \& {The Goods Team}}]{Popesso09}
{Popesso}, P., {Dickinson}, M., {Nonino}, M., {et~al.} 2009, \aap, 494, 443

\bibitem[{{Raiter} {et~al.}(2010){Raiter}, {Fosbury}, \&
  {Teimoorinia}}]{Raiter2010}
{Raiter}, A., {Fosbury}, R.~A.~E., \& {Teimoorinia}, H. 2010, \aap, 510, A109+

\bibitem[{{Razoumov} \& {Sommer-Larsen}(2009)}]{Razoumov09}
{Razoumov}, A.~O. \& {Sommer-Larsen}, J. 2009, ArXiv e-prints

\bibitem[{{Scalo}(1986)}]{Scalo86}
{Scalo}, J.~M. 1986, Fundamentals of Cosmic Physics, 11, 1

\bibitem[{{Schaerer}(2002)}]{Scha02}
{Schaerer}, D. 2002, \aap, 382, 28

\bibitem[{{Schaerer}(2003)}]{Scha03}
{Schaerer}, D. 2003, \aap, 397, 527

\bibitem[{{Schaerer}(2008)}]{Schaerer08}
{Schaerer}, D. 2008, in IAU Symposium, Vol. 255, IAU Symposium, ed.
  {L.~K.~Hunt, S.~Madden, \& R.~Schneider}, 66--74

\bibitem[{{Schaerer} \& {de Barros}(2010)}]{SdB10}
{Schaerer}, D. \& {de Barros}, S. 2010, \aap, 515, A73+

\bibitem[{{Schaerer} \& {Pell{\'o}}(2005)}]{Schaerer2005}
{Schaerer}, D. \& {Pell{\'o}}, R. 2005, \mnras, 362, 1054

\bibitem[{{Schaerer} \& {Vacca}(1998)}]{SV98}
{Schaerer}, D. \& {Vacca}, W.~D. 1998, \apj, 497, 618

\bibitem[{{Schaerer} \& {Verhamme}(2008)}]{SV07}
{Schaerer}, D. \& {Verhamme}, A. 2008, \aap, 480, 369

\bibitem[{{Schneider} {et~al.}(2002){Schneider}, {Ferrara}, {Natarajan}, \&
  {Omukai}}]{schne02}
{Schneider}, R., {Ferrara}, A., {Natarajan}, P., \& {Omukai}, K. 2002, \apj,
  571, 30

\bibitem[{{Schneider} {et~al.}(2003){Schneider}, {Ferrara}, {Salvaterra},
  {Omukai}, \& {Bromm}}]{schne03}
{Schneider}, R., {Ferrara}, A., {Salvaterra}, R., {Omukai}, K., \& {Bromm}, V.
  2003, \nat, 422, 869

\bibitem[{{Stasi{\'n}ska} \& {Schaerer}(1999)}]{Stasinska99}
{Stasi{\'n}ska}, G. \& {Schaerer}, D. 1999, \aap, 351, 72

\bibitem[{{Stasi{\'n}ska} \& {Tylenda}(1986)}]{Stasinska86}
{Stasi{\'n}ska}, G. \& {Tylenda}, R. 1986, \aap, 155, 137

\bibitem[{{Stiavelli}(2009)}]{sti09}
{Stiavelli}, M. 2009, {From First Light to Reionization: The End of the Dark
  Ages}, ed. {Stiavelli, M.}

\bibitem[{{Tumlinson}(2006)}]{Tumlinson06}
{Tumlinson}, J. 2006, \apj, 641, 1

\bibitem[{{Tumlinson} {et~al.}(2001){Tumlinson}, {Giroux}, \&
  {Shull}}]{Tumlinson01}
{Tumlinson}, J., {Giroux}, M.~L., \& {Shull}, J.~M. 2001, \apjl, 550, L1

\bibitem[{{Tumlinson} \& {Shull}(2000)}]{Tumlinson00}
{Tumlinson}, J. \& {Shull}, J.~M. 2000, \apjl, 528, L65

\bibitem[{{Vanzella} {et~al.}(2008){Vanzella}, {Cristiani}, {Dickinson},
  {Giavalisco}, {Kuntschner}, {Haase}, {Nonino}, {Rosati}, {Cesarsky},
  {Ferguson}, {Fosbury}, {Grazian}, {Moustakas}, {Rettura}, {Popesso},
  {Renzini}, {Stern}, \& {The Goods Team}}]{Vanzella08}
{Vanzella}, E., {Cristiani}, S., {Dickinson}, M., {et~al.} 2008, \aap, 478, 83

\bibitem[{{Vanzella} {et~al.}(2005){Vanzella}, {Cristiani}, {Dickinson},
  {Kuntschner}, {Moustakas}, {Nonino}, {Rosati}, {Stern}, {Cesarsky}, {Ettori},
  {Ferguson}, {Fosbury}, {Giavalisco}, {Haase}, {Renzini}, {Rettura}, {Serra},
  \& {The Goods Team}}]{Vanzella05}
{Vanzella}, E., {Cristiani}, S., {Dickinson}, M., {et~al.} 2005, \aap, 434, 53

\bibitem[{{Vanzella} {et~al.}(2006){Vanzella}, {Cristiani}, {Dickinson},
  {Kuntschner}, {Nonino}, {Rettura}, {Rosati}, {Vernet}, {Cesarsky},
  {Ferguson}, {Fosbury}, {Giavalisco}, {Grazian}, {Haase}, {Moustakas},
  {Popesso}, {Renzini}, {Stern}, \& {The Goods Team}}]{Vanzella06}
{Vanzella}, E., {Cristiani}, S., {Dickinson}, M., {et~al.} 2006, \aap, 454, 423

\bibitem[{{Vanzella} {et~al.}(2009){Vanzella}, {Giavalisco}, {Dickinson},
  {Cristiani}, {Nonino}, {Kuntschner}, {Popesso}, {Rosati}, {Renzini}, {Stern},
  {Cesarsky}, {Ferguson}, \& {Fosbury}}]{Vanzella09}
{Vanzella}, E., {Giavalisco}, M., {Dickinson}, M., {et~al.} 2009, \apj, 695,
  1163

\bibitem[{{Wise} \& {Cen}(2009)}]{Wise09}
{Wise}, J.~H. \& {Cen}, R. 2009, \apj, 693, 984

\bibitem[{{Yamada} {et~al.}(2005){Yamada}, {Sasaki}, {Sumiya}, {Umeda},
  {Shioya}, {Ajiki}, {Nagao}, {Murayama}, \& {Taniguchi}}]{Yamada05}
{Yamada}, S.~F., {Sasaki}, S.~S., {Sumiya}, R., {et~al.} 2005, \pasj, 57, 881

\end{thebibliography}

\end{document}